\documentclass[12pt,draftcls,peerreview,twocolumn]{IEEEtran}

\renewcommand{\baselinestretch}{1.65}
\usepackage{amsmath, graphics,amssymb,epsfig,subfigure,color}

\newtheorem{theorem}{Theorem}

\newtheorem{lemma}{Lemma}

\begin{document}

\title{{ Multi-Way Information Exchange Over Completely-Connected Interference Networks with a Multi-Antenna Relay}
}

\author{
  \IEEEauthorblockN{Namyoon Lee and Robert W. Heath Jr.}\\
  \IEEEauthorblockA{ Wireless Networking and Communications Group\\
 Department of Electrical and Computer Engineering
\\  The University of Texas at
Austin, Austin, TX 78712 USA\\
    E-mail~:~namyoon.lee@utexas.edu and rheath@utexas.edu}
}



\maketitle

\begin{abstract}
This paper considers a fully-connected interference network with a relay in which multiple users equipped with a single antenna want to exchange multiple unicast messages with other users in the network by sharing the relay equipped with multiple antennas. For such a network, the degrees of freedom (DoF) are derived by considering various message exchange scenarios: a multi-user fully-connected Y channel, a two-pair two-way interference channel with the relay, and a two-pair two-way X channel with the relay. Further, considering distributed relays employing a single antenna in the two-way interference channel and the three-user fully-connected Y channel, achievable sum-DoF are also derived in the two-way interference channel and the three-user fully-connected Y channel. A major implication of the derived DoF results is that a relay with multiple antennas or multiple relays employing a single antenna increases the capacity scaling law of the multi-user interference network when multiple directional information flows are considered, even if the networks are \textit{fully-connected} and all nodes operate in \textit{half-duplex}. These results reveal that the relay is \textit{useful} in the multi-way interference network with practical considerations.
\end{abstract}


\section{Introduction}


Multi-way communication using an intermediate relay is a promising wireless network architecture with applications including cellular networks, sensor networks, and device-to-device communications. The simplest multi-way relay network model is the two-way relay channel \cite{Wu}-\cite{Rankov} in which a pair
of two users wish to exchange messages by sharing
a single relay. Although the capacity of this simple channel
is still unknown in general \cite{Nam}, physical layer network coding \cite{Larsson}-\cite{Popovski}  and analog network coding \cite{Katti}-\cite{Rankov} increase the achievable sum-rates of the two-way relay channels because it allows users to exploit their transmit signal as side-information. Recently, the two-way relay channel has been generalized in a number of ways:  multi-pair two-way relay channels \cite{Chen:09}-\cite{Sezgin} and multi-user multi-way relay channels \cite{Gunduz:09}-\cite{Lee_Chun:11}. For the multi-pair two-way relay channel where multiple user pairs exchange messages with their partners by sharing a common relay, the authors \cite{Sezgin} characterized the capacity of multi-pair two-way relay network for a deterministic and Gaussian
channel model. For the multi-user multi-way relay channel with unicast messages exchange setup, the multiple-input multiple-output (MIMO) Y channel was introduced where three users exchange independent unicast messages with each other via an intermediate relay and characterized the degrees of freedom (DoF) of the channel by the idea of \textit{signal space alignment for network coding}  \cite{Lee_Lim_Chun:10}. This result was extended into the case of a general number of users as a $K$-user Y channel \cite{Lee_Lee_Lee:12}.

In spite of extensive studies on different multi-way relay channels, relatively little work has been addressed on the understanding of the capacity of multi-way relay channel, especially when the nodes are fully-connected in the network due to difficulty in managing interference. Note that when the direct links are considered in the multi-way relay channel, it can be equivalently viewed as an interference network with a relay. In general, if the networks are fully-connected, i.e., they have a non-layered structure, then a node receives signals arriving along different paths, which causes more inter-user interference than that of the layered network. When the interference networks have a layered structure, it has been shown that the relay can offer gain in the number of DoF \cite{Gou_Jafar_Jeon_Chung}-\cite{Shomorony}. On the other hand, for the fully-connected interference network with relays \cite{Tannious_Nosratinia:08}-\cite{Tian}, it was shown that relays \textit{cannot} not improve the DoF of such a network regardless of how many antennas the relay has with a few exceptions where the cognitive relay \cite{S_S_S:2008} or the instantaneous relay \cite{Lee_Wang:12} is considered. If the non-layered multi-hop interference network supports uni-directional information flows \cite{Shomorony_Avestimehr:11}-\cite{Wang_Gou_Jafar_layeredX}, the non-layered structure incurs a DoF loss. Even for multi-way information flows, relays with infinitely many antennas \textit{do not} increase the DoF of the fully-connected X network whose source nodes are disjoint from destination nodes \cite{Cadambe_Jafar:2009} .
In this paper, we provide counter examples of the claim that the relay cannot increase the DoF of the fully-connected interference network. We show that the relay is useful in improving the DoF of the multi-user interference networks when multi-way information exchange is allowed between users.

In this paper, we consider a fully-connected interference network with a relay where $K$ users with a single antenna exchange unicast messages with each other via a relay with $N$ multiple antennas. In particular, we assume that all nodes have half-duplex  constraint due to hardware limitations, implying that transmission and reception occurs in different orthogonal time slots. We consider three different multi-way information exchange setups over the fully-connected interference network.\begin{itemize}

\item \textbf {Fully-connected Y channel with a multi-antenna relay}: First, we consider $K$-user fully-connected interference network with a relay equipped with $N$ antennas. In such a channel, each user wishes to send $K-1$ unicast messages to all other users and also wishes to decode all
other users' messages. Since the same message exchange setup is considered in the previous work on Y channel \cite{Lee_Lim_Chun:10} and \cite{Lee_Lee_Lee:12} without direct links between users, we refer to it as ``\textit{a fully-connected Y channel}."

\item  \textbf {Two-pair two-way interference channel a multi-antenna relay}: As a special case of the four-user fully-connected Y channel, we consider a four-user fully-connected interference network with a relay where the four users form two-pairs. The two pairs exchange messages with their partners via a multi-antenna relay. In particular, when the relay node is ignored, this channel model is equivalent with the two-way interference channels studied in \cite{Suh} and \cite{Devroye}.

\item  \textbf {Two-way X channel with a multi-antenna relay}: We also consider another four-user completely-connected  interference network with a two-antenna relay. In this channel, each user wants to exchange two unicast messages with two different users in the network. Since this channel model can be viewed as a bi-directional X channel, we refer to it as \textit{``two-way X channel with a relay.''}

\end{itemize}

The main contribution of this paper is to derive sum-DoF bounds for certain networked channels. Specifically, for the general multi-way information exchange case, i.e., a fully-connected Y channel, it is demonstrated that the sum-DoF of $\frac{K(K-1)}{2K-2}$ is the optimal if the relay has $N\geq K-1$ antennas by showing both converse and achievability. Further, it is shown that the sum-DoF of $\frac{4}{3}$ and $\frac{8}{5}$ are achievable for the two-pair two-way interference and X channel with a relay when the relay has $N=2$. These result are interesting because it has been shown that the sum-DoF for the fully-connected interference network cannot be improved by the use of the relays even if the relay has infinitely many antennas \cite{Chen_Cheng:10} and \cite{Cadambe_Jafar:2009}. Our result, however, reveals the fact that a relay with a finite number of antennas can increase the DoF of the network when multi-directional communication is considered between pairs. As an extension, by considering multiple distributed relays which of each has a single antenna, we derive the sum-DoF bounds for both the two-pair two-way interference and three-user fully-connected Y channel. One major implication of the results is that the available DoF of fully-connected interference network that supports multi-directional information exchange can be improved substantially by allowing a relay with multiple antennas or  the multiple distributed relays even if all nodes operate in \textit{half-duplex}. To see how the relay is useful in terms of DoF for the multi-way communications, it is instructive to compare our result with the case when no relay is used by using the following examples:

\begin{itemize}

\item Example 1 (Four-user fully-connected Y channel without a relay): Let us consider a four-user, fully-connected, and half-duplex Y channel where each user wants to exchange three unicast messages with the other users in the network. If we assume that there is no relay in the network, the optimal DoF of such a channel equals $\frac{4}{3}$ as shown in \cite{Cadambe_Jafar:2009}, which can be achieved by interference alignment. Meanwhile, our result shows that the sum-DoF of $2$ is achievable by involving a relay employing $N=3$ antennas, which is a $50\%$ DoF improvement.

\item Example 2 (Two-pairs two-way interference channel without a relay) : When the relay is not considered in the two-pairs two-way interference channel, it is known that the optimal DoF is one \cite{Cadambe_Jafar:08}. By involving the relay with two antennas in the network, however, it is possible to achieve the sum-DoF of $\frac{4}{3}$, which is $33\%$ DoF increase.
\end{itemize}


Why does the relay provide DoF gain for multi-way communication ? The DoF gain comes from two mechanisms. One is the side-information inherently given by multi-way communication, i.e., \textit{caching gain}. The second is due to the fact that the relay can make sure each user does not see the undesired interference signal or
see the same interference shape using joint space-time precoding techniques. We refer to it as \textit{interference shaping gain}. To acquire two different gains, the multiple antenna relay or multiple relays with a single antenna controls the information flow of the multi-way communication so that each user exploits side-information efficiently, which leads to increase the DoF by the use of a relay. To show our results, multi-phase transmission schemes  are proposed, which are inspired by wireless index coding \cite{Birk} and \cite{Namyoon}.

%
%
The rest of the paper is organized as follows. In Section II,
a general system model of multi-way communications with a MIMO relay is described. To provide the intuition behind the proposed transmission schemes, a motivating example is provided in Section III.  In Section IV, the optimal DoF for the fully-connected Y channel is addressed. Section V provides the sum-DoF inner bounds for the two-pair two-way interference and X channel with a relay by considering different message exchange scenarios in the four-user fully-connected interference network. To see the effect of relay antenna cooperation, achievable sum-DoF bounds are derived for the two-pair two-way interference channel and 3-user fully-connected Y channel when three distributed relays are considered in Section VI. The paper is concluded in Section VII.

Throughout this paper, transpose, conjugate transpose, inverse, Frobenius norm and
trace of a matrix ${\bf X}$ are represented by ${\bf X}^{T}$ ,
${\bf X}^{*}$, ${\bf X}^{-1}$, $\|{\bf X}\|_F$  and $\textmd{Tr}\left({\bf
X}\right)$, respectively. In addition, $\mathbb{C}$ and $\mathbb{R}$
indicates a complex and real value. $\mathcal{CN}(0,1)$
represents a complex Gaussian random variable with zero mean and unit variance.

\section{System Model}
\begin{figure}
\centering
\includegraphics[width=4in]{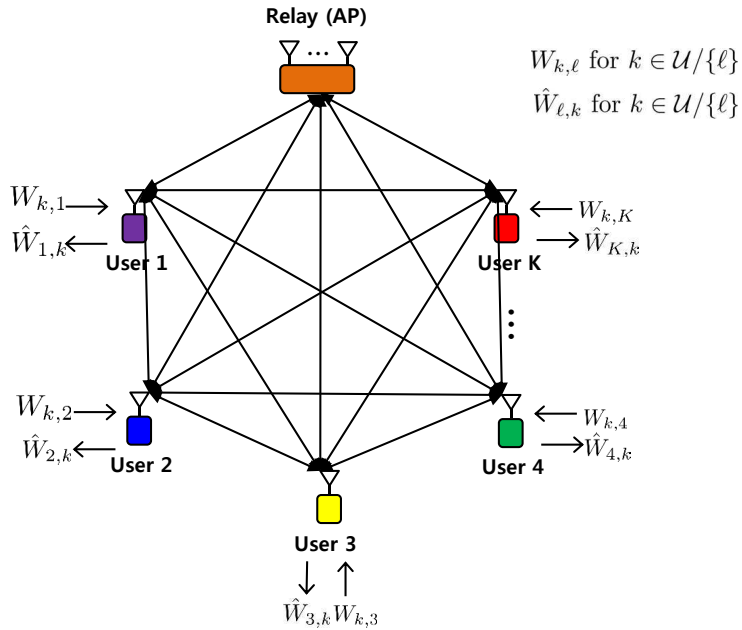}
\caption{A $K$-user fully-connected interference network with one relay. User $\ell$ wants to send $K-1$ messages $W_{k,\ell}$ and receive ${\hat W}_{\ell,k}$ for $k\in \mathcal{U}/\{\ell\}$ in this network. } \label{fig:1}
\end{figure}

Let us consider an interference network comprised of $K$ users with a single antenna each and a relay with $N$ antennas. All the users and the relay are completely-connected as illustrated in Fig. \ref{fig:1}. User $k$, $k\in\mathcal{U}\triangleq \{1,2,\ldots,K\}$, wants to send $K-1$ unicast messages $W_{\ell,k}$ for $\ell\in \mathcal{U}/\{k\}\triangleq \mathcal{U}_k^{\textrm{c}}$ to user $\ell$ and intends to decode $K-1$ messages $W_{k,\ell}$ for $\ell \in \mathcal{U}_k^{\textrm{c}}$ sent by all other users. In this channel, it is assumed that the relay and all nodes operate in half-duplex mode, implying that transmission and reception span orthogonal time slots.

Let $x_{\ell}[n]=f(W_{k,\ell})$ for $k\in \mathcal{U}_{\ell}^{c}$ be the transmitted signal by user $\ell$ at time slot $n$ where $f(\cdot)$ represents an encoding function. Also, let $\mathcal{S}_n$ and $\mathcal{D}_n$ denote the set of source and destination nodes at time slot $n$. Due to the fully-connected property and the half-duplex constraint, when the users belonging the source set $\mathcal{S}_n$ send their signals at the $n$-th time slot simultaneously, user $k\in \mathcal{D}_n$ and the relay receives the signals
\begin{eqnarray}
{y}_{k}[n] &=& \sum_{\ell \in \mathcal{S}_n}{h}_{k,\ell}[n]x_{\ell}[n] + {z}_{k}[n], \quad  k\in\mathcal{D}_n,\\
{\bf y}_{\textrm{R}}[n] &=& \sum_{\ell \in \mathcal{S}_n}{\bf h}_{\textrm{R},\ell}[n]x_{\ell}[n] + {\bf z}_{\textrm{R}}[n],
\end{eqnarray}
where $y_k[n]$ and ${\bf y}_{\textrm{R}}[n]\in \mathbb{C}^{N \times 1} $ represent the received signal at user $k$ and the relay;
${z}_k[n]$ and ${\bf z}_R[n]$ denote the additive noise signal at user $k$ and at the relay at time slot $n$ whose elements
are Gaussian random variable with zero mean and unit variance,
i.e., $\mathcal{CN}(0,1)$: and ${{h}_{k,\ell}}[n]$ and ${{\bf h}_{\textrm{R},\ell}}[n]=[ {h}^1_{\textrm{R},\ell}[n], \ldots,  {h}^N_{\textrm{R},\ell}[n]]$ represent the channel coefficients from user $\ell$ to user $k$ and the channel vector from user $\ell$ to the relay, respectively.

When the relay and user $\ell \in S_n$ cooperatively transmit at the $n$-th time slot, at the same time, user $k\in \mathcal{D}_n$ receives the signal as
\begin{eqnarray}
{ y}_{k}[n]&=& \sum_{\ell \in \mathcal{S}_n}{h}_{k,\ell}[n]x_{\ell}[n]+ {\bf h}^*_{k,\textrm{R}}[n]{\bf x}_{\textrm{R}}[n]+ {z}_{k}[n],\quad k\in\mathcal{D}_n,
\end{eqnarray}
where ${\bf h}^*_{j,\textrm{R}}[n]=[ {h}^1_{j,\textrm{R}}[n], \ldots,  {h}^N_{j,\textrm{R}}[n]]^*$ denotes the (downlink) channel vector from the relay to user $k$ and ${\bf x}_{\textrm{R}}[n]$ represents the transmit signal vector at the relay when the $n$-th channel is used.

The transmit power at each user and the relay is assumed to be $P$,
i.e., $\mathbb{E}\left[|{x}_{j}[n]|^2\right] \leq P$ and
$\mathbb{E}\left[\|{\bf x}_{\textrm{R}}[n]\|_2^2\right] \leq P$. Further, it is assumed that all the entries of all
channel values in ${h}_{\ell,k}[n]$, ${\bf h}_{\textrm{R},\ell}[n]$, and ${\bf h}^*_{k,\textrm{R}}[n]$ are drawn from a continuous
distribution and the absolute value of all the channel coefficients is bounded between a nonzero minimum value and a finite maximum value. The channel state information (CSI) is assumed to be perfectly known at the users in receiving mode and the relay has global channel knowledge for all links.

User $k$ sends an independent message  $W_{\ell,k}$ for one intended user $\ell$ with rate $R_{\ell,k}(P)=\frac{\log_2|W_{\ell,k}|}{n}$ for $\ell, k \in\mathcal{U}$ and $\ell \neq k$, a rate tuple
$\mathcal{R}=\left(R_{1,2},\textrm{R}_{1,3},\ldots,
R_{K,K-1}\right)\in{\mathbb{R}^{K(K-1)}}$ is achievable if every receiver can
decode the desired message with an error probability that is
arbitrarily small with sufficient channel uses $n$. Then, the sum-DoF characterizing the approximate sum-rates in the
high \textrm{SNR} regime is defined as
\begin{align}
d_{\textrm{sum}}&=\sum_{k=1,k\neq \ell}^{K}\sum_{\ell=1}^{K}d_{k,\ell} \nonumber \\
&=\lim_{{P}\rightarrow \infty}\frac{\sum_{k=1,k\neq \ell}^{K}\sum_{\ell=1}^{K}R_{k,\ell}\left({P}\right)}{\log\left({P}\right)}.
\end{align}
In this work, the sum-DoF is a key metric to compare the network performance of different message setups.

\section{Motivating Example}
Before deriving the main results, in this section, we provide an example that yield intuition about why the relay is useful for increasing the DoF in a three-user fully-connected Y channel.

\subsection{Proposed Multi-Phase Transmissions}
\begin{figure}
\centering
\includegraphics[width=5in]{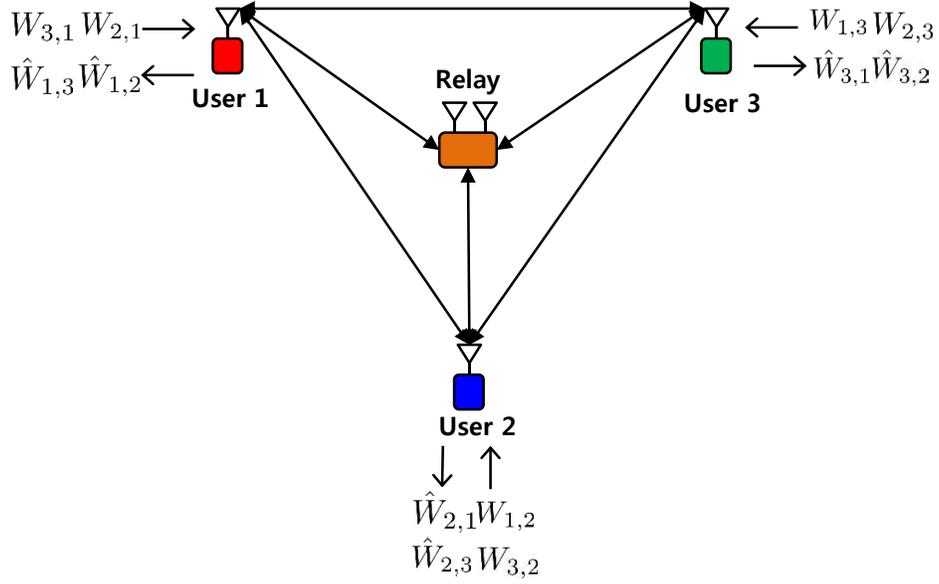}
\caption{Three-user fully-connected Y channel. } \label{fig:example2}
\end{figure}

Consider a 3-user fully-connected Y channel with a $N=2$ antennas relay. As illustrated in Fig.\ref{fig:example2}, in this channel each user sends two independent messages, one to each other user. Since there are direct links between users, this network model differs from previous work on the Y channel \cite{Lee_Lim_Chun:10} and \cite{Chaaban} where the direct links between users are ignored. We will show that $\frac{3}{2}$ sum-DoF is achievable.

\subsubsection{Phase One (Round-Robin Multiple-Access Channel (MAC) )} This phase comprises of three time slots. In each channel use, two users send one message to one intended user so that the intended user has one equation that contains two desired data symbols. Specifically,
at time slot 1, user 2 and 3 send information symbols $s_{1,2}$ and $s_{1,3}$ for user 1. While user 1 and the relay listen the signals, user 2 and 3 do not receive any signals in this time slot due to half-duplex constraint, i.e. $\mathcal{S}_1=\{2,3\}$ and $\mathcal{D}_1=\{1\}$.  When noise is ignored, user 1 and the relay have\begin{align}
D_1[1]&=h_{1,2}[1]s_{1,2} + h_{1,3}[1]s_{1,3}, \\
{\bf D}_R[1]&={\bf h}_{\textrm{R},2}[1]s_{1,2} + {\bf h}_{\textrm{R},3}[1]s_{1,3}.
\end{align}
Since the relay has two antennas, it resolves the transmitted data symbols $s_{1,2}$ and $s_{1,3}$ by using a zero-forcing (ZF) decoder.

In the second time slot, user 1 and user 3 transmit data symbols $s_{2,1}$ and $s_{2,3}$ to user 2. The received equations at  user 1 and the relay are
 \begin{align}
D_2[2]&=h_{2,1}[2]s_{2,1} + h_{2,3}[2]s_{2,3}, \\
{\bf D}_R[2]&={\bf h}_{\textrm{R},1}[2]s_{2,1} + {\bf h}_{\textrm{R},3}[2]s_{2,3}.
\end{align}
By taking an advantage of multiple antennas, the relay decodes $s_{2,1}$ and $s_{2,3}$.

Finally, at time slot 3, user 1 and user 2 deliver data symbols $s_{3,1}$ and $s_{3,2}$ to user 3. Hence, the received equation at  user 3 and the relay are given by
\begin{align}
D_3[3]&=h_{3,1}[3]s_{3,1} + h_{3,2}[3]s_{3,2}, \\
{\bf D}_R[3]&={\bf h}_{\textrm{R},1}[3]s_{3,1} + {\bf h}_{\textrm{R},2}[3]s_{3,2}.
\end{align}
Similarly, the relay obtains $s_{3,1}$ and $s_{3,2}$ by using a ZF decoder. As a result, during the phase one, each user obtains one equation consisted of two desired symbols and the relay acquires all six independent data symbols in the network.

\subsubsection{Phase Two (Relay Broadcast)}
The second phase spans one time slot. In this time slot, the relay sends a superposition of six data symbols obtained during the phase one. The transmitted signal at the relay is given by
\begin{eqnarray}
{\bf x}_R[4]= \sum_{i=1}^{3}\sum_{j=1, j\neq i}^{3}{\bf v}_{i,j}[4]s_{i,j},
\end{eqnarray}
where ${\bf v}_{i,j}[4]\in \mathbb{C}^{2\times 1}$ denotes the
beamforming vector used for carrying symbol $s_{i,j}$ at time slot $4$. The main design principle of ${\bf v}_{i,j}[4]$ is to control the interference propagation on the network so that each user receives an equation that consists of desired data symbols or self interference data symbols which can be eliminated by using side-information at each user. For instance, user 1 wants to receive an additional equation consisted of $s_{1,2}$ and $s_{1,3}$ and can cancel the self-interference signals caused by $s_{2,1}$ and $s_{3,1}$ by exploiting side-information. Thus, the relay picks the beamforming vectors ${\bf v}_{2,3}[4]$ and ${\bf v}_{3,2}[4]$ carrying $s_{2,3}$ and $s_{3,2}$ so that user 1 does not receive them. To accomplish this,  ${\bf v}_{2,3}[4]$ and ${\bf v}_{3,2}[4]$ are selected as
\begin{eqnarray}
{\bf v}_{2,3}[4] \in \textrm{null}({\bf h}_{1,\textrm{R}}^*[4])~ \textrm{and} ~   {\bf v}_{3,2}[4] \in \textrm{null}({\bf h}_{1,\textrm{R}}^*[4]).
\end{eqnarray}
Applying the same principle, the other relay beamforming vectors are designed as
\begin{align}
&{\bf v}_{1,3}[4] \in \textrm{null}({\bf h}_{2,\textrm{R}}^*[4]), ~  {\bf v}_{3,1}[4] \in \textrm{null}({\bf h}_{2,\textrm{R}}^*[4]),  \\
&{\bf v}_{1,2}[4] \in \textrm{null}({\bf h}_{3,\textrm{R}}^*[4]), ~ \textrm{and} ~  {\bf v}_{2,1}[4] \in \textrm{null}({\bf h}_{3,\textrm{R}}^*[4]).
\end{align}
To give some intuition on the proposed precoding solution, we rewrite the transmit signal at the relay as
\begin{eqnarray}
{\bf x}_R[4]&=& {\bf v}^{c}_1[4](s_{2,3}+s_{3,2})+{\bf v}^{c}_2[4](s_{1,3}+s_{3,1}) +{\bf v}^{c}_3[4](s_{1,2}+s_{2,1}),
\end{eqnarray}
where ${\bf v}^{c}_i[4] \in \textrm{null}({\bf h}_{i,\textrm{R}}^*[4])$. Thus, we can interpret the transmitted signal at the relay at the second phase as a class of \textit{superposition coding}.

 \subsubsection{Decoding}
We explain a decoding method used by user 1. Recall that user 1 received an equation consisting of two desired symbols $s_{1,2}$ and $s_{1,3}$ at time slot 1 in the form of $D_1[1]=h_{1,2}[1]s_{1,2}+h_{1,3}[1]s_{1,3}$. In time slot 4, user 1 obtained an equation containing both two desired and two self-interference data symbols given by
\begin{align}
y_{1}[4] &= {\bf h}^{*}_{1,\textrm{R}}[4]{\bf x}_R[4],\nonumber \\
&={\bf h}^{*}_{1,\textrm{R}}[4]\left\{{\bf v}^{c}_2[4](s_{1,3}\!+\!s_{3,1})\!+\!{\bf v}^{c}_3[4](s_{1,2}\!+\!s_{2,1})\right\}.
\end{align}
Assuming that user 1 preserves the transmitted information symbols $s_{2,1}$ and $s_{3,1}$ by caching memory as side-information and it has the effective channel ${\bf h}^{*}_{1,\textrm{R}}[4]{\bf v}^{c}_i[4]$ for $i\in\{1,2,3\}$, user 1 can generate an interference equation $M_1[4]= {\bf h}^{*}_{1,\textrm{R}}[4]{\bf v}^{c}_2[4]s_{3,1} + {\bf h}^{*}_{1,\textrm{R}}[4]{\bf v}^{c}_3[4]s_{2,1} $. Thus, user 1 extracts one equation that contains two desired symbols by eliminating the self-interference as
\begin{align}
&y_{1}[4] - M_1[4]={\bf h}^{*}_{1,\textrm{R}}[4]{\bf v}^{c}_3[4]s_{1,2}+ {\bf h}^{*}_{1,\textrm{R}}[4]{\bf v}^{c}_2[4]s_{1,3}.
\end{align}
From self-interference cancellation, we acquired a new equation for the two desired symbols. Therefore, the two desired symbols are obtained by solving the following matrix equation,
\begin{eqnarray}
\left[\!\!\!%
\begin{array}{c}
  {y}_1[1] \\
  {y}_1[4]-M_{1}[4] \\
\end{array}%
\!\!\!\right]\!=\!\underbrace{\left[\!\!%
\begin{array}{cc}
  h_{1,2}[1] & h_{1,3}[1] \\
 {\bf h}^{*}_{1,\textrm{R}}[4]{\bf v}^{c}_3[4] &{\bf h}^{*}_{1,\textrm{R}}[4]{\bf v}^{c}_2[4]\\
\end{array}\!\!%
\right]}_{{\bf \tilde{H}}_{1}}\!\!\!\left[\!\!%
\begin{array}{c}
  s_{1,2} \\
  s_{1,3} \\
\end{array}\!\!%
\!\right]\!.
\end{eqnarray}
Note that beamforming vectors were selected independently of the direct channel between users, i.e. $h_{1,2}[1]$ and $h_{1,3}[1]$. Therefore, the rank of the effective channel becomes 2 with probability one, which allows user 1 to decode two desired symbols $s_{1,2}$ and $s_{1,3}$. For user 2 and user 3, the same method applies. Consequently, it is possible to exchange a total of six data symbols within 4 channel uses by using the relay employing multiple antennas in the 3-user fully-connected Y channel.



\subsection{Interpretation of the Proposed Methods}

Now we reinterpret our results from the perspective of index coding.
The basic index coding problem is a follows. Suppose a transmitter has a set of information messages $W=\{W_1,W_2,\ldots,W_K\}$ for multiple receivers and each receiver wishes to receive a subset of $W$ while knowing some other subset of $W$ as side information. The underlying goal of index coding is to design the best encoding strategy at the transmitter using the side-information at the receivers to minimize the number of transmissions, while allowing all receivers to obtain their desired messages.

The proposed transmission schemes mimic the index coding algorithms developed in \cite{Birk}-\cite{Namyoon}. Specifically, until the relay has global knowledge of messages in the network,  $N$ users propagate information into the network at each time slot. Since the relay has $N$ antennas, it obtains $N$ information symbols per one time slot and the remaining $K-N$ other users in receiving mode acquires one equation that has both desired and interfering symbols. When the relay obtains all the messages, it starts to control information flow by sending a useful signal to all users so that each user decodes the desired information symbols efficiently based on their previous knowledge: their transmitted symbols and the received equations. For instance, there are six messages $\{W_{2,1},W_{3,1},W_{1,2},W_{3,2},W_{1,3},W_{2,3}\}$ in the three-user Y channel. During time slot 1, 2 and 3, the relay acquires global message set $\{W_{2,1},W_{3,1},W_{1,2},W_{3,2},W_{1,3},W_{2,3}\}$  and each user acquires following side information
\begin{itemize}
	\item User 1 knows $\{W_{2,1},W_{3,1}\}$ and $L_1(W_{1,2},W_{1,3})$,
	 \item User 2 knows $\{W_{1,2},W_{3,2}\}$ and $L_2(W_{2,1},W_{2,3})$,
	\item User 3 knows $\{W_{1,3},W_{2,3}\}$ and $L_{3}(W_{3,1},W_{3,2})$,
\end{itemize}
where $L_k(A,B)$ denotes a linear function of the two messages $A$ and $B$ obtained at user $k$. In time slot 4, the relay broadcasts the six mixed independent data symbols so that each user may decode their desired data symbols by using the message they sent (caching) and the overheard interference signals in the previous phases. This relay transmission allows for the users  to obtain gains for caching and interference shaping gains in the network.
Thus, we can interpret this relay precoding solution as a linear vector index coding in a complex field for a special class of index coding problems.

\section{DoF of $K$-User Fully-Connected Y Channel  }
In this section, we consider a general setup where each user wants to exchange $K-1$ independent messages with all other users in the network as described in Section II. For such a channel, the following theorem is the main result of this section.

\begin{theorem} \label{Theorem1}
For the fully-connected Y channel where $K$ users have a single antenna and a relay has $N \geq K-1$ antennas, the maximum sum-DoF equals $\frac{K}{2}$.
\end{theorem}

\begin{figure}
\centering
\includegraphics[width=5in]{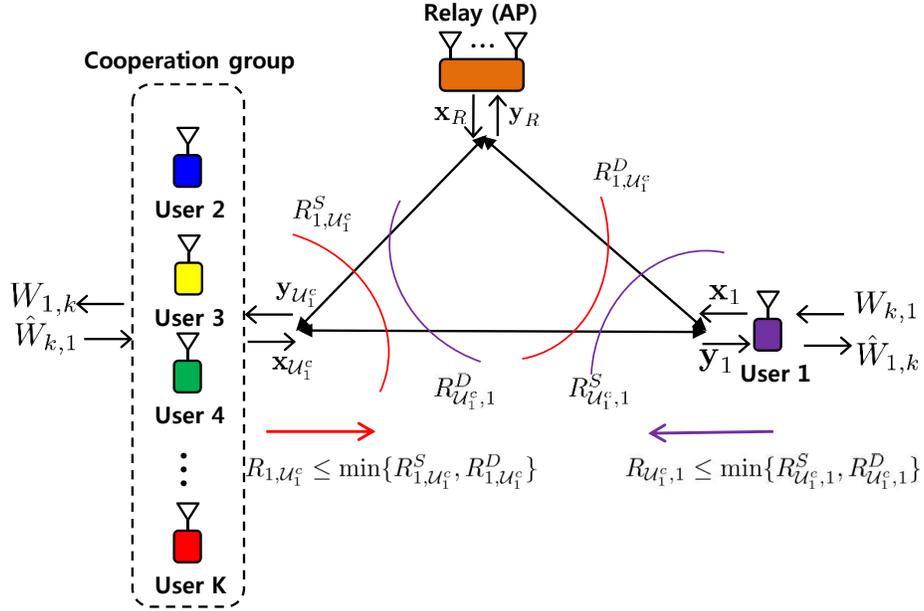}
\caption{A user cooperation scenario when all users except for user 1 cooperate by sharing the messages and antennas. Thus, $K$-user fully-connected Y channel is equivalently viewed as a non-separate two-way relay channel where one node has $K-1$ antennas but the other node has a single antenna. In this cooperation case, the messages between cooperating users set to be null, i.e., $W_{i,j}=\phi$  for $i,j\in \mathcal{U}_1^{\textrm{c}}$. For this setup, cut-set bounds are applied.} \label{fig:upper}
\end{figure}

\subsection{Converse}
We provide the converse of Theorem \ref{Theorem1} by using the cut-set theorem.
Using the fact that user cooperation does not deteriorate the DoF of the channel, let us first consider a special case where all users except user $k$ fully cooperate. This cooperation allows us to view the fully-connected Y channel as a two-way relay channel equivalently where the user group has $K-1$ antennas but user 1 has a single antenna as illustrated in Fig \ref{fig:upper}. Further, let us set the messages to be null between cooperating users, i.e., $W_{i,j}=\phi$  for $i,j\in \mathcal{U}_1^{\textrm{c}}$, by the using the fact that the null-messages cannot degrade the performance of the non-null messages. In this two-way relay channel, the user group wants to send the message $W_{1,k}$ for $k\in\mathcal{U}_1^{\textrm{c}}$ and user 1 wants to send the messages $W_{k,1}$ for $k\in\mathcal{U}_1^{\textrm{c}}$. The converse follows from the following lemma which serves an outer bound of the equivalent two-way relay channel.

\begin{lemma} \label{Lemma1} Let $d_{k,\ell}$ be the DoF for message $W_{k,\ell}$ for $k, \ell \in \mathcal{U}$. Then, the following inequality holds:
\begin{align}
\sum_{\ell=1,\ell\neq k}^{K}d_{k,\ell} + \sum_{k=1,k\neq\ell}^{K}d_{k,\ell} \leq 1, \quad \textrm{for} \quad k,\ell \in\mathcal{U}.
\end{align}
\end{lemma}
\textit{Proof:} The detailed proof is provided in Appendix A. Note that for a single antenna at all nodes, an outer bound of the non-separate two-way relay channel is derived in \cite{outter}. We extend this outer bound result to the case where the relay and a user have multiple antennas.

To attain the converse result of Theorem \ref{Theorem1}, we add $K$ inequalities from Lemma 1, which gives us
\begin{align}
2\left(\sum_{ \ell\neq k}^{K}\sum_{k=1}^{K}d_{\ell,k}\right) \leq K \\
\Rightarrow \sum_{ \ell\neq k}^{K}\sum_{k=1}^{K}d_{\ell,k} \leq \frac{K}{2},
\end{align}
which completes the proof.

\subsection{Achievability}
Communication takes place in two phases: 1) the multiple access (MAC) phase where $K-1$ users send an independent message to one intended user and the relay overhears the signal; 2) a relay broadcast phase where
the relay transmits the signal obtained during the previous phase to all the users.

\subsubsection{Phase One (MAC phase)} The first phase comprises of $K$ time slots, i.e., $T_1=\{1,2,\ldots,K\}$. For $k\in T_1$ time slot, user $k$ and the relay listen the transmitted signals by all other users, i.e., $\mathcal{S}_k =\mathcal{U}_{k}^{\textrm{c}}$ and $\mathcal{D}_k=\{k\}$. Specifically, at time slot $k$ all the users in $\mathcal{U}_{k}^{\textrm{c}}$ sends information symbol $\{s_{k,\ell}\}$ for $k\neq \ell$ to user $k$ simultaneously. Thus, the received signals at user $k$ and the relay are given by
\begin{align}
{y}_{k}[k] &= \sum_{\ell=1,\ell\neq k}^{K}{h}_{k,\ell}[k]s_{k,\ell} + {z}_{k}[k], \label{eq:received11}\\
{\bf y}_{\textrm{R}}[k] &= \sum_{\ell=1,\ell\neq k}^{K}{\bf h}_{\textrm{R},\ell}[k]s_{k,\ell} + {\bf z}_{\textrm{R}}[k],
\end{align}
Note that user $k$ acquires a linear equation consisting of the desired symbols. Meanwhile, the relay can decode $K-1$ information symbols from the users by using a ZF decoder since
it has $N\geq K-1$ antennas. As a result, during phase 1, each user obtains one desired equation and the relay acquires global knowledge of the $K(K-1)$ messages on the network.

\subsubsection{Phase Two (Relay Broadcast)}
The second phase spans $K-2$ time slots, $T_2=\{K+1,\ldots,2K-1\}$. In each time slot, the relay sends a superposition of $K(K-1)$ data symbols obtained during the phase one. The transmitted signal at the relay is given by
\begin{eqnarray}
{\bf x}_R[n]= \sum_{\ell \neq k}^{K}\sum_{k=1}^{K}{\bf v}_{k,\ell}[n]s_{\ell,k}, \quad n\in T_2,
\end{eqnarray}
where ${\bf v}_{\ell,k}[n]\in \mathbb{C}^{N\times 1}$ denotes the
precoding vector used for carrying symbol $s_{\ell,k}$ at time slot $n $. The principle for desinging beamforming vectors ${\bf v}_{\ell,k}[n]$ is to control the information flow so that each user receives an equation consisting of desired data symbols and known interference. For instance, user 1 wants to receive an additional equation that consists of symbols $\{s_{1,2},~s_{1,3}, \ldots, ~s_{1,K}\}$ and has the capability to remove the interference terms cased by its transmitted symbols $\{s_{2,1},~s_{3,1}~,
\ldots,s_{K,1}\}$ by exploiting caching memory. Using this fact, the relay picks the precoding vector ${\bf v}_{\ell,k}[n]$ carrying $s_{\ell,k}$ so that user $j$ for $j\in \mathcal{U}/\{k,\ell\}$ does not receive it. To accomplish this,  ${\bf v}_{\ell,k}[n]$ is selected as
\begin{eqnarray}
{\bf v}_{\ell,k}[n] \in \textrm{null}\left(\underbrace{\left[%
\begin{array}{c}
  {\bf h}_{i_1,\textrm{R}}^*[n] \\
  {\bf h}_{i_2,\textrm{R}}^*[n] \\
  \vdots \\
  {\bf h}_{i_{K-2},\textrm{R}}^*[n] \\
\end{array}%
\right]}_{(K-2)\times N}\right),
\end{eqnarray}
where $\{ i_1, i_2, \ldots, i_{K-2}\} = \mathcal{U}/\{\ell,k\}$ denotes an index set with $K-2$ elements. Note that the relay has $N\geq K-1$ antennas and the elements of channel vectors are drawn from i.i.d. random variables, the precoding solution of ${\bf v}_{\ell,k}[n]$ exists with probability one. When the relay transmits during the second phase, the received signal at user $j$ is given by
\begin{align}
y_{j}[n] &={\bf h}^{*}_{j,\textrm{R}}[n]{\bf x}_R[n] \nonumber \\
&= {\bf h}^{*}_{j,\textrm{R}}[n]\sum_{\ell \neq k}^{K}\sum_{k=1}^{K}{\bf v}_{k,\ell}[n]s_{\ell,k}
\nonumber \\
&=\underbrace{{\bf h}^{*}_{j,\textrm{R}}[n]\sum_{k=1,k\neq j}^{K}{\bf v}_{j,k}[n]s_{j,k}}_{D_j[n]}+\underbrace{{\bf h}^{*}_{j,\textrm{R}}[n]\sum_{i=1,i\neq j}^{K}{\bf v}_{i,j}[n]s_{i,j}}_{M_j[n]} +z_{j}[n]. \label{eq:rx_y}
\end{align}
In (\ref{eq:rx_y}), $D_j[n]$ contains the $K-1$ desired symbols seen by user $j$ at time slot $n\in T_2$ and $M_j[n]$ consists of known symbols transmitted by user $j$.

\subsubsection{Decoding}
Let us consider a decoding for user $j$. Recall that user $j$ received an equation consisting of $K-1$ desired symbols $\{s_{j,1},\ldots s_{j,j-1},s_{j,j+1},\ldots ,s_{j,K}\}$ during time slot $j\in T_1$, i.e., $D_j[j]=\sum_{k=1,k\neq j}^{K}h_{j,k}[j]s_{j,k}$. Further, for time slots that belong to the second phase $n\in T_2$, it acquired $K-2$ additional linear equations each of which contains both $K-1$ desired and $K-1$ known data symbols
\begin{align}
y_{j}[n] &=  D_j[n] + M_j[n]+ z_{j}[n].
\end{align}
First, the receiver subtracts the known interference term by exploiting knowledge of side-information $\{s_{1,j},\ldots s_{j-1,j},s_{j+1,j},\ldots ,s_{K,j}\}$. Assuming that user $j$ knows the effective downlink channel ${\bf h}_{j,\textrm{R}}^*[n]{\bf v}_{k,j}[n]$ for $k\in \mathcal{U}/\{j\}$, the receiver generates the same interference shape $M_j[n]$. After interference cancellation, the remaining equation contains the desired $K-1$ desired symbols
\begin{align}
{ \tilde y}_j[n]&=y_{j}[n] - M_j[n]=D_j[n] + z_j[n], \quad n\in T_2 \label{eq:received22}
\end{align}
Finally, to decode $K-1$ intended information symbols, the equations in (\ref{eq:received11}) and (\ref{eq:received22}) are aggregated into a matrix form,
\begin{eqnarray}
\left[\!\!\!%
\begin{array}{c}
  {y}_j[j] \\
  {\tilde y}_j[K\!+\!1]\\
  \vdots \\
  {\tilde y}_j[2K\!-\!2] \\
\end{array}\!\!\!%
\right]\!\!\!=\!\underbrace{\left[\!\!\!\!%
\begin{array}{ccccccc}
  h_{j,1}[j] \!&\! \cdots \!\!\!&\!\!\!  h_{j,j-1}[j] \!\!\!&\!\!\! h_{j,j+1}[j] \!\!\!&\!\!\! \cdots \!\!\!&\!\!\! h_{j,K}[j]\\
  {\tilde h}_{j,1}[K\!+\!1] \!\!\!&\!\!\! \cdots \!\!\!&\!\!\!  {\tilde h}_{j,j-1}[K\!+\!1] \!\!&\!\! {\tilde h}_{j,j+1}[K\!+\!1] \!\!&\!\! \cdots \!\!&\!\! {\tilde h}_{j,K}[K\!+\!1]\\
\vdots & \ddots & \vdots & \vdots  &\ddots & \vdots\\
  {\tilde h}_{j,1}[2K\!-\!2] \!&\! \cdots &  {\tilde h}_{j,j-1}[2K\!-\!2] & {\tilde h}_{j,j+1}[2K\!-\!2] & \cdots & {\tilde h}_{j,K}[2K\!-\!2]\\
\end{array}\!\!\!\!%
\right]}_{{\bf \tilde{H}}_{j}}\!\!\!\left[\!\!%
\begin{array}{c}
  s_{j,1} \\
\vdots \\
  s_{j,j-1} \\
  s_{j,j+1} \\
\vdots \\
s_{j,K} \\
\end{array}\!\!%
\right],
\end{eqnarray}
where ${\tilde h}_{j,k}[n]$ denotes the effective channel coefficient from the relay to user $j$ carrying information symbol $s_{j,k}$ at time slot $n\in T_2$, i.e., ${\tilde h}_{j,k}[n]={\bf h}^{*}_{j,\textrm{R}}[n]{\bf v}_{j,k}[n]$. It is important to note that the beamforming vectors were selected statistically independent with respect to the direct channel between users, i.e. $h_{j,k}[j]$ for $k \in\mathcal{U}_{j}^{\textrm{c}}$ at time slot $j\in T_1$. Therefore, the effective channel matrix ${\bf \tilde{H}}_{j}$ has full rank, i.e., $\textrm{rank}({\bf \tilde{H}}_{j})=K-1$ with probability one, which allows user $j$ to decode $K-1$ desired symbols $s_{j,k}$ for $k \in \mathcal{U}_{j}^{\textrm{c}}$ at user $j$. By symmetry, user $k$ for $k \in \mathcal{U}_{j}^{\textrm{c}}$ can apply the same decoding method to obtain $K-1$ desired data symbols. As a result, it is possible to exchange a total of $K(K-1)$ data symbols within $K+K-2$ channel uses by using the relay in the fully-connected Y channel, which leads to achieve $\frac{K(K-1)}{2K-2}=\frac{K}{2}$ sum-DoF. This completes the proof.

Now, we make several remarks on the implication of our results.

\textit{Remark 1 ({No CSIT at users}):} To achieve the optimal DoF, while CSIT at users is not needed, the users require to know the effective (downlink) channel value from the relay to user $j$ for $j\in \mathcal{U}$, i.e., ${\tilde h}_{j,k}[n]={\bf h}^{*}_{j,\textrm{R}}[n]{\bf v}_{j,k}[n]$ for performing self-interference cancellation. This channel knowledge can be obtained using demodulation reference signals that currently used in commercial wideband systems. Alternatively, CSIT at the relay plays in important role to attain the DoF gains.

\textit{Remark 2 ({Decoding delay}):} Note that each user cannot decode the desired information symbols until the relay transmissions are finished, which results in decoding delay. To reduce the delay, the proposed strategy may be implemented in a multi-carrier system that offers $K-2$ independent parallel sub-channels. From this system, the relay can send all information symbols required at all users for decoding within one time slot but over $K-2$ independent sub-channels.


\textit{Remark 3 ({Full-duplex operation}):} In this work, we assumed that all nodes are in half-duplex. One interesting observation is that the DoF increase by the use of a relay shown this work can also be translated to the gain overcoming half-duplex loss. For example, consider a $K$-user fully-connected X network, which is the same as the fully-connected Y channel without the relay. In this network, if the half-duplex is assumed, the optimal DoF is $d_{\textrm{half}}^X=\frac{(K/2)(K/2 -1)}{K-1}$ as shown in \cite{Cadambe_Jafar:2009}. Meanwhile, for the full-duplex assumption, the lower and upper bounds of the optimal DoF were $\frac{K}{2} \leq d^X_{\textrm{full}} \leq \frac{K(K-1)}{2K-3} $ \cite{Cadambe_Jafar:2009}. Since our result assuming the half-duplex operation meets the tight inner bound DoF for the full-duplex X network, i.e., $\frac{K}{2}$, the relay allows users to overcome the loss due to half-duplex signaling in the network.

\section{Four-User Fully-Connected Interference Network with Different Messages}

In this section we consider two examples for the four-user fully-connected interference network. Specifically, we derive DoF inner bounds for two different channel models: two-pair two-way interference channel with a relay and two-pair two-way X channel with a relay.

\subsection{Example 1: Two-Pair Two-Way Interference Channel with a Relay}  
\begin{figure}
\centering
\includegraphics[width=4in]{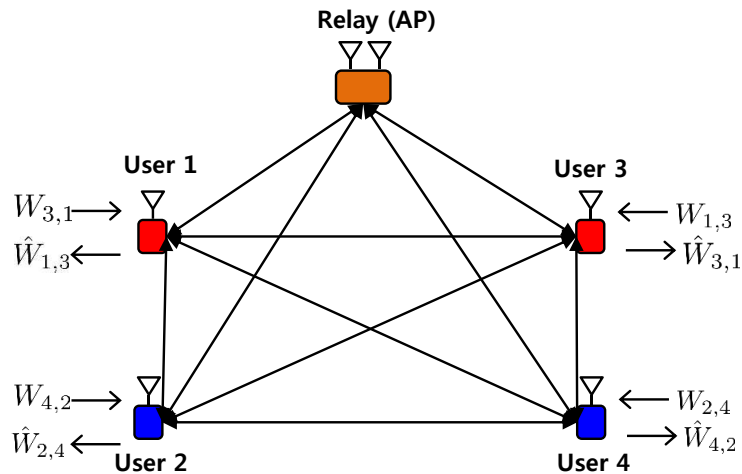}
\caption{The two-pair two-way interference channel with two antennas relay. Each user wants to exchange the messages with its partner by using a shared relay.} \label{fig:2}
\end{figure}

In this example, we assume that there are four users in the network, i.e., $K=4$ and the relay has two antennas $N=2$ as depicted in Fig \ref{fig:2}. Among the four users,  two user pairs, user 1-user 3 and user 2-user 4, want to exchange the messages. Since each user exchanges the message with its partner in a bi-directional way, we refer to it as the two-pair two-way interference channel with a relay. Note that this channel is equivalent to the four-user fully-connected interference network with a relay when the eight messages are set to be null, i.e., $W_{2,1}=W_{4,1}=\phi$, $W_{1,2}=W_{3,2}=\phi$, $W_{2,3}=W_{4,3}=\phi$, and $W_{1,4}=W_{3,4}=\phi$. Throughout this example, we show that four independent symbols $s_{1,3} =f(W_{1,3})$, $s_{3,1}=f(W_{3,1})$, $s_{2,4}=f(W_{2,4})$, and $s_{4,2}=f(W_{4,2})$ can be exchanged over three time slots using a new multi-phase transmission method, which allows users to exploit side-information efficiently.

%

\subsubsection{Phase 1 (Forward Interference Channel (IC) Transmission)} Phase one consists of one time slot. In this phase, user 1 and user 2 transmit signals $x_1[1]=s_{3,1}$ and $x_2[1]=s_{4,2}$ over the forward IC, i.e., $\mathcal{S}_{1}=\{1,2\}$ and $\mathcal{D}_{1}=\{3,4\}$. Note that user 1 and 2 cannot receive each other's signals during time slot 1 due to the half-duplex constraint. When noise is neglected, the received signals at user 3, user 4, and the relay are given by
\begin{eqnarray}
{ y}_{3}[1] &=&{h}_{3,1}[1]s_{3,1}+  {h}_{3,2}[1]s_{4,2}, \label{eq:example1} \\
{ y}_{4}[1] &=& {h}_{4,1}[1]s_{3,1}+  {h}_{4,2}[1]s_{4,2}, \label{eq:example2}\\
{\bf y}_{\textrm{R}}[1]&=&{\bf h}_{\textrm{R},1}[1]s_{3,1} +{\bf h}_{\textrm{R},2}[1] s_{4,2}, \label{eq:example3}
\end{eqnarray}
Since the relay has two antennas, it resolves the transmitted data symbols $s_{3,1}$ and $s_{4,2}$ by using a zero-forcing (ZF) decoder\footnote{One may use another spatial decoders that can resolve two independent symbols at the relay with two antennas such as minimum mean square error (MMSE) or Vertical-Bell Laboratories Layered Space-Time (V-BLAST). For simplicity, ZF decoder will be used in this paper. }.
\subsubsection{Phase 2 (Backward IC Transmission)} Phase 2 uses one time slot. In time slot 2, information flow occurs over the backward channel, i.e., $\mathcal{S}_2=\{3,4\}$ and $\mathcal{D}_2=\{1,2\}$. Specifically, user $3$ and user $4$ send signals ${x}_{3}[2]={ s}_{1,3}$ and ${ x}_{4}[2]={s}_{{2,4}}$ at time slot 2 through the backward IC. The signal received at the receivers of the backward channel are given by
\begin{eqnarray}
y_1[2]&=&{ h}_{1,3}[2]{ s}_{1,3}+{ h}_{1,4}[2]{ s}_{{2,4}} \nonumber \\
y_2[2]&=&{ h}_{2,3}[2]{ s}_{{1,3}}+{ h}_{2,4}[2]{ s}_{{2,4}}\nonumber \\
{\bf y}_{\textrm{R}}[2]&=&{\bf { h}}_{\textrm{R},3}[2]{ s}_{1,3} +{\bf { h}}_{\textrm{R},4}[2] { s}_{{2,4}}.
\end{eqnarray}
From the backward transmission, user 1 and user 2 obtain a linear equation, while the relay decodes ${ s}_{{1,3}} $ and ${ s}_{2,4} $  using a ZF decoder.

\subsubsection{Phase 3 (The Relay Transmission) } Phase 3 also uses one time slot. The real novelty occurs in this time slot using a technique that inspired by a wireless index coding. In this time slot, the signal transmitted by the relay is
 \begin{eqnarray}
{\bf x}_{\textrm{R}}[3]={\bf v}_{3,1}[3]s_{3,1}+{\bf {v}}_{1,3}[3]{ s}_{1,3}+{\bf v}_{4,2}[3]s_{4,2}+{\bf {v}}_{2,4}[3]{ s}_{2,4}, \label{eq:DF}
 \end{eqnarray}
where ${\bf v}_{i,j}[3]$ denotes the precoding vector for information symbol $s_{i,j}$ for $i,j\in\{1,2,3,4\}$. The relay transmission aims to multicast the signal ${\bf x}_R[3]$ so that all users can decode the desired information symbol based on the side-information each user acquired from the previous time slots.  For example, user 1 wants to decode data symbol ${ s}_{1,3}$. Two different forms of side-information are acquired: the transmitted symbol ${s}_{3,1}$ at time slot 1 and the received signal from the backward transmission at time slot 2, i.e., $y_1[2]={ h}_{1,3}[2]{ s}_{1,3}+{ h}_{1,4}[2]{ s}_{{2,4}}$. To exploit this side-information when user 1 decodes the desired symbols efficiently, the relay should not propagate the interference symbol $s_{4,2}$ to user 1. Hence, we select the relay precoding vector ${\bf v}_{4,2}[3]$ carrying information symbol $s_{4,2}$ in a such that it does not reach to user 1 by selecting
 \begin{eqnarray}
{\bf v}_{4,2}[3] \in \textrm{null}\left({\bf h}^*_{1,\textrm{R}}[3]\right). \label{eq:example_relay1} \end{eqnarray}
To accomplish the same objective for the other users, we choose the relay precoding vectors to satisfy
 \begin{eqnarray}
{\bf v}_{3,1}[3]&\in& \textrm{null}\left({\bf h}^*_{2,\textrm{R}}[3]\right),\label{eq:example_relay2}  \\
{\bf {v}}_{2,4}[3] &\in& \textrm{null}\left({\bf {h}}^*_{3,\textrm{R}}[3]\right), \label{eq:example_relay3} \\
{\bf {v}}_{1,3}[3]&\in& \textrm{null}\left({\bf  {h}}^*_{4,\textrm{R}}[3]\right).\label{eq:example_relay4}
 \end{eqnarray}
Since the size of the channels ${\bf h}^*_{k,\textrm{R}}[3]$ for $k\in\mathcal{U}$ is $1\times 2$, the beamforming solutions satisfying the equations in (\ref{eq:example_relay1}), (\ref{eq:example_relay2}), (\ref{eq:example_relay3}), and (\ref{eq:example_relay4}) exist almost surely. Thus, the received signals at users at time slot 3 are given by
\begin{eqnarray}
{y}_{1}[3]&=&{\bf h}^*_{1,\textrm{R}}[3]{\bf x}_{\textrm{R}}[3] \nonumber \\
&=&{\bf h}^*_{1,\textrm{R}}[3]{\bf v}_{1,3}[3]s_{1,3}+{\bf h}^*_{1,\textrm{R}}[3]{\bf {v}}_{2,4}[3]{s}_{2,4}+\underbrace{{\bf h}^*_{1,\textrm{R}}[3]{\bf { v}}_{3,1}[3]{ s}_{3,1}}_{\textrm{Self-interference}},
\end{eqnarray}
\begin{eqnarray}
{y}_{2}[3]&=&{\bf h}^*_{2,\textrm{R}}[3]{\bf x}_{\textrm{R}}[3] \nonumber \\
&=&{\bf h}^*_{2,\textrm{R}}[3]{\bf v}_{1,3}[3]s_{1,3}+{\bf h}^*_{2,\textrm{R}}[3]{\bf {v}}_{2,4}[3]{ s}_{2,4}+\underbrace{{\bf h}^*_{2,\textrm{R}}[3]{\bf { v}}_{4,2}[3]{ s}_{4,2}}_{\textrm{Self-interference}},\end{eqnarray}
\begin{eqnarray}
{{y}}_{3}[3]&=&{\bf {h}}^*_{3,\textrm{R}}[3]{\bf x}_{\textrm{R}}[3], \nonumber \\
&=&{\bf {h}}^*_{3,\textrm{R}}[3]{\bf v}_{3,1}[3]s_{3,1}+{\bf h}^*_{3,\textrm{R}}[3]{\bf v}_{4,2}[3]s_{4,2}+\underbrace{{\bf {h}}^*_{3,\textrm{R}}[3]{\bf v}_{1,3}[3]{s}_{1,3}}_{\textrm{Self-interference}},
\end{eqnarray}
\begin{eqnarray}
{y}_{4}[3]&=&{\bf {h}}^*_{4,\textrm{R}}[3]{\bf x}_{\textrm{R}}[3],\nonumber \\
&=&{\bf h}^*_{4,\textrm{R}}[3]{\bf v}_{3,1}[3]s_{3,1}+{\bf h}^*_{4,\textrm{R}}[3]{\bf v}_{4,2}[3]s_{4,2}+\underbrace{{\bf h}^*_{4,\textrm{R}}[3]{\bf {v}}_{2,4}[3]{s}_{2,4}}_{\textrm{Self-interference}}.
\end{eqnarray}

\subsubsection{Decoding} Successive interference cancellation is used to eliminate the back propagating self-interference from the received signal at time slot 3. The remaining inter-user interference is removed by a ZF decoder. For instance, user $1$ eliminates the self-interference ${\bf h}^*_{1,\textrm{R}}[3]{\bf { v}}_{3,1}[3]{ s}_{3,1}$ from $y_1[3]$ as
 \begin{align}
{y}_{1}[3]-{\bf h}^*_{1,\textrm{R}}[3]{\bf { v}}_{3,1}[3]{ s}_{3,1}={\bf h}^*_{1,\textrm{R}}[3]{\bf v}_{1,3}[3]s_{1,3}+{\bf h}^*_{1,\textrm{R}}[3]{\bf {v}}_{2,4}[3]{s}_{2,4}.
\end{align}
After canceling the self-interference, the received signals at time slot 2 and time slot 3 can be rewritten in matrix form as
\begin{eqnarray}
\left[%
\begin{array}{c}
  {y}_{1}[2]\\
 {y}_{1}[3]-{\bf h}^*_{1,\textrm{R}}[3]{\bf { v}}_{3,1}[3]{ s}_{3,1}\\
\end{array}%
\right]=\underbrace{\left[%
\begin{array}{cc}
 {h}_{1,3}[2] & { h}_{1,4}[2] \\
{\bf h}^*_{1,\textrm{R}}[3]{\bf {v}}_{1,3}[3] & {\bf h}^*_{1,\textrm{R}}[3]{\bf v}_{2,4}[3]\\
\end{array}%
\right]}_{{\bf \tilde{H}}_{1}}\left[%
\begin{array}{c}
  { s}_{1,3} \\
 { s}_{2,4} \\
\end{array}%
\right] .
\end{eqnarray}
Since the beamforming vectors ${\bf {v}}_{1,3}[3]$ and ${\bf { v}}_{2,4}[3] $ were designed independently of $ {\bf h}^*_{1,\textrm{R}}[3]$ and the channel coefficients ${h}_{1,3}[2] $ and ${h}_{1,4}[2] $ were drawn from a continuous random distribution, the effective channel matrix ${\bf \tilde{H}}_{1}$ has full rank almost surely. This implies that it is possible to decode the desired symbol ${ s}_{1,3}$ by applying a ZF decoder that eliminates the effect of inter-user interference ${ s}_{2,4}$. Consequently, user 1 obtains the desired data symbol ${ s}_{{1,3}}$. By symmetry, the other users are able to decode the desired symbols by using the same decoding procedure. As a result, a total $4$ of independent data symbols are delivered over three orthogonal channel uses, which leads to achieve the $\frac{4}{3}$ of sum-DoF, i.e., $d_{\textrm{sum}}=\frac{4}{3}$.

\textit{Remark 4 (CSI knowledge and feedback)}: To cancel interference, it is assumed that each user has knowledge of the effective channel from the relay to the users, i.e., $\{ {\bf h}^*_{k,\textrm{R}}[3]{\bf v}_{1,3}[3], {\bf h}^*_{k,\textrm{R}}[3]{\bf v}_{3,1}[3], {\bf h}^*_{k,\textrm{R}}[3]{\bf v}_{4,2}[3], {\bf h}^*_{k,\textrm{R}}[3]{\bf v}_{2,4}[3]\}$ for $k\in \mathcal{U}$. This effective channel, however, can be estimated using demodulation reference signals formed in commercial wideband systems. With the setting, the users do not need to know CSIT, implying that no CSI feedback is required between users. In contrast, the relay needs to know CSIT between the relay to the users to generate precoding vectors. While this CSIT can by given by a feedback link if frequency division duplexing system is considered, it can be acquired without feedback when time division duplex system is applied due to channel reciprocity.

\textit{Remark 5 (Another Transmission Method)}: It is possible to achieve the $\frac{4}{3}$ of sum-DoF by applying a different transmission scheme. During time slot 1, user 1 and user 3 send signals and the other users and the relay listen, i.e., $\mathcal{S}_1=\{1,3\}$ and $\mathcal{D}_1=\{2,4\}$. At time slot 2,  while user 2 and user 4 send the signals, other nodes overhear the transmitted signals, i.e., $\mathcal{S}_2=\{2,4\}$ and $\mathcal{D}_2=\{1,3\}$. In the third time slot, the relay can perform space-time interference alignment \cite{Namyoon}, which creates the same interference shape between the currently observed and the previously acquired interference signal.
Thus, each user cancels the received interference at time slot 3 from the acquired interference equations during time slot 1 and 2. The details of the scheme is included in Appendix B.

\textit{Remark 6 ({Amplify-and-Forward (AF) vs Decode-and-Forward (DF)}):} In this example, we assumed that the relay uses DF relaying. The same DoF ca be achieved by AF relaying. Let ${\bf U}[n]$ denotes the spatial decoder used at the relay for $n\in\{1,2\}$. When AF relaying is considered, the transmitted signal at time slot 3 in (\ref{eq:DF}) is modified as
 \begin{eqnarray}
{\bf x}_{\textrm{R}}[3]=[{\bf v}_{3,1}[3],~{\bf v}_{4,2}[3]]{\bf U}[1]{\bf y}_{\textrm{R}}[1]+[{\bf v}_{1,3}[3],~{\bf v}_{2,4}[3]]{\bf U}[2]{\bf y}_{\textrm{R}}[2].
 \end{eqnarray}

\subsection{Example 2: Two-Pair Two-Way X Channel with a Relay}

\begin{figure}
\centering
\includegraphics[width=5in]{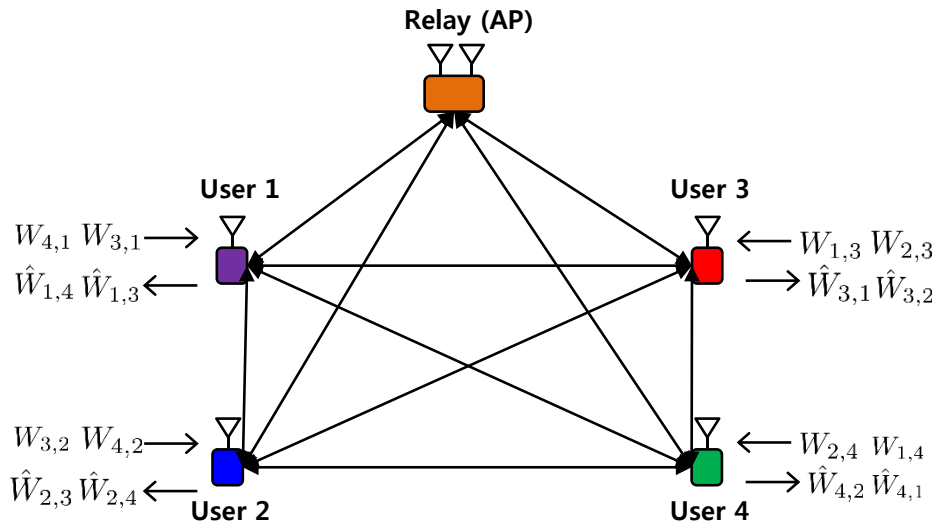}
\caption{The two-pair two-way X channel with two antennas relay. Each user wants to exchange two independent messages with the other user group by using a shared relay.} \label{fig:system}
\end{figure}

In this example, we assume that $K=4$, and $N=2$. Although the physical channel model is the same as Example 1, we consider a more complex information exchange scenario where user 1 and 2 want to exchange two independent messages with both user 3 and 4. We refer to this scenario as a two-pair two-way X channel with a multiple antenna relay. Note that this channel can be interpreted as a 4-user multi-way interference network in which $W_{2,1}=W_{1,2}=\phi$ and $W_{3,4}=W_{4,3}=\phi$. In this example, we will show that each user exchanges two independent symbols with two different users over five time slots;
a total $\frac{8}{5}$ sum-DoF are achievable. This achievability is shown by a generalization of space-time interference alignment \cite{Namyoon}.

\subsubsection{Phase One (Forward IC transmission)} This phase consists of two time slots.
In the first time slot, user 1 and user 2 send an independent
symbol intended for user $3$, i.e., ${x}_{1}[1]=s_{3,1}$ and
${x}_{2}[1]=s_{3,2}$. In the second time slot, user 1 and user 2 transmit independent
symbols intended for user $4$, i.e., ${x}_{1}[2]=s_{4,1}$ and
${x}_{2}[2]=s_{4,2}$, i.e., $\mathcal{S}_n=\{1,2\}$ and $\mathcal{D}_n=\{3,4\}$ for $n\in\{1,2\}$. Let denote $D_{j}[n]$ and $L_j[n]$ denote the received equations at user $j$ in the $n$-th time slot, which contain the desired and interference symbols, respectively. Neglecting noise at the receivers, user $3$ and user $4$ obtain two linear equations during two time slots, which are
\begin{eqnarray}
{ D}_{3}[1] &=&{h}_{3,1}[1]s_{3,1}+  {h}_{3,2}[1]s_{3,2},   \\
{ L}_{4}[1] &=&{h}_{4,1}[1]s_{3,1}+  {h}_{4,2}[1]s_{3,2},  \\
{ L}_{3}[2] &=& {h}_{3,1}[2]s_{4,1}+  {h}_{3,2}[2]s_{4,2}, \\
{ D}_{4}[2] &=& {h}_{4,1}[2]s_{4,1}+  {h}_{4,2}[2]s_{4,2}.
\end{eqnarray}
Due to the broadcast nature of the wireless medium, the relay is also able to listen the transmissions by the users. Since it has two antennas, it is possible to decode two information symbols in each part of phase one, giving $s_{3,1}$, $s_{3,2}$, $s_{4,1}$, and $s_{4,2}$, by using a ZF decoder during the phase one.

\subsubsection{Phase Two  (Backward IC Transmission)} In the second phase, the role of transmitters and receivers is reversed, i.e., $\mathcal{S}_n=\{3,4\}$ and $\mathcal{D}_n=\{1,2\}$ for $n\in\{3,4\}$. In time slot 3, user $3$ and user $4$ send an independent
symbol intended for user 1, ${ {x}}_{1}[3]={ s}_{1,3}$ and
${{x}}_{2}[3]={ s}_{1,4}$. For time slot 4, user $3$ and user $4$ deliver information symbols intended for user 2, ${x}_{3}[4]= {s}_{2,3}$ and ${x}_{4}[4]={ s}_{2,4}$. Therefore, user 1 and user 2 obtain two equations during the phase two, which are given by

\begin{eqnarray}
{ D}_{1}[3] &=&{h}_{1,3}[3]s_{1,3}+  {h}_{1,4}[3]s_{1,4},
 \\
{ L}_{2}[3] &=&{h}_{2,3}[3]s_{1,3}+  {h}_{2,4}[3]s_{1,4},  \\
{ L}_{1}[4] &=& {h}_{1,3}[4]s_{2,3}+  {h}_{1,4}[4]s_{2,4}, \\
{ D}_{2}[4] &=& {h}_{2,4}[4]s_{2,3}+  {h}_{2,4}[4]s_{2,4} ,  
\end{eqnarray}
As with phase two, the relay decodes four data symbols ${ s}_{1,3}$, ${ s}_{1,4}$, ${ s}_{2,3}$, and ${s}_{2,4}$ by using a ZF decoder.

\subsubsection{Phase Three (Relay Broadcast)} The third phase uses only one time slot. In this phase, in contrast to previous examples, the relay exploits knowledge of the current downlink CSI from the relay to the users, i.e., ${\bf h}_{k,\textrm{R}}[5]$ for $k \in \{1,2,3,4\}$ and outdated CSI between users i.e., $\left\{h_{4,1}[1],h_{4,2}[1],h_{3,1}[2],h_{3,1}[2]\right\}$ as well as the outdated CSI between the users, i.e., $\left\{{ h}_{2,3}[3],{h}_{2,4}[3],{ h}_{1,3}[3],{h}_{1,4}[4] \right\}$. Using this information, in time slot 5, the relay transmit the eight data symbols $\{s_{3,1},~s_{3,2},~s_{4,1},~s_{4,2},~{s}_{3,1},~{s}_{3,2},~{s}_{4,1},~{ s}_{4,2}\}$ acquired during the phase one and two  as
\begin{eqnarray}
{\bf x}_{\textrm{R}}[5]= \sum_{j=3}^{4}\sum_{i=1}^{2}{\bf v}_{j,i}[5]{s}_{j,i} +  \sum_{j=1}^{2}\sum_{i=3}^{4}{\bf { v}}_{j,i}[5]{{ s}}_{j,i},
\end{eqnarray}
where ${\bf v}_{j,i}[5]\in \mathbb{C}^{2\times 1}$ denotes the
beamforming vector to carry symbol ${s}_{i,j}$ during time slot 5. The main idea of the relay beamforming design is to provide the same interference signal shape to all users observed through phase one and phase two so that each user can use the received interference signal during phase two as side information.

To illustrate, we explain the design principle of ${\bf v}_{3,1}[5]$ carrying $s_{3,1}$ from a index coding perspective. Note that  data symbol $s_{3,1}$ is only desired by user $3$ and it is interference to all the other users except for user 1. This is because user 1 has already $s_{3,1}$, so it can use it as side-information for self-interference cancellation. User 4 observed $s_{3,1}$ at time slot 1 in the form of ${L}_{4}[1]=h_{4,1}[1]s_{3,1}+h_{4,2}[1]s_{3,2}$. Therefore, user $4$ can cancel $s_{3,1}$ from the relay transmission if it receives the same interference shape $h_{4,1}[1]s_{3,1}$.  Unlike user 4, user 2 does not have any knowledge of $s_{3,1}$. Thus, the relay must design the beamforming vector carrying $s_{3,1}$ so that it does not reach to user 2. To satisfy both user 2 and 4, the relay designs  ${\bf v}_{3,1}[5]$  as
\begin{align}
{\bf h}_{2,\textrm{R}}^*[5]{\bf v}_{3,1}[5]&=0,  \nonumber \\
{\bf h}_{4,\textrm{R}}^*[5]{\bf v}_{3,1}[5]&= h_{4,1}[1].
\end{align}
By applying the same design principle, we pick the other precoding vectors so that the following conditions are satisfied as
\begin{align}
\left[%
\begin{array}{c}
{{\bf h}}_{1,\textrm{R}}^*[5]\\
{{\bf h}}_{3,\textrm{R}}^*[5] \\
\end{array}%
\right]{\bf v}_{4,1}[5]=
\left[
     \begin{array}{c}
           0\\
	h_{3,1}[2]\\
         \end{array}
       \right], \quad
\left[%
\begin{array}{c}
{{\bf h}}_{1,\textrm{R}}^*[5]\\
{{\bf  h}_{4,\textrm{R}}^*}[5] \\
\end{array}%
\right]{\bf v}_{3,2}[5]=
\left[
         \begin{array}{c}
           0\\
	h_{4,2}[1]\\
         \end{array}
       \right],
\end{align}
\begin{align}
\left[%
\begin{array}{c}
{{\bf h}}_{1,\textrm{R}}^*[5]\\
{{\bf h}_{3,\textrm{R}}^*}[5] \\
\end{array}%
\right]{\bf v}_{4,2}[5]=
\left[
         \begin{array}{c}
           0\\
	h_{3,2}[2]\\
         \end{array}
       \right], \quad
\left[%
\begin{array}{c}
{{\bf { h}}_{4,\textrm{R}}^*}[5] \\
{{\bf h}}_{2,\textrm{R}}^*[5]\\
\end{array}%
\right]{\bf {v}}_{1,3}[5]=
\left[
     \begin{array}{c}
           0\\
	{ h}_{2,3}[3]\\
         \end{array}
       \right],
\end{align}
\begin{align}
\left[%
\begin{array}{c}
{{\bf { h}}_{4,\textrm{R}}^*}[5] \\
{{\bf h}}_{1,\textrm{R}}^*[5]\\
\end{array}%
\right]{\bf {v}}_{2,3}[5]=
\left[
     \begin{array}{c}
           0\\
	{ h}_{1,3}[4]\\
         \end{array}
       \right], \quad
\left[%
\begin{array}{c}
{{\bf { h}}_{3,\textrm{R}}^*}[5] \\
{{\bf h}}_{2,\textrm{R}}^*[5]\\
\end{array}%
\right]{\bf {v}}_{1,4}[5]=
\left[
         \begin{array}{c}
           0\\
	{ h}_{2,4}[3]\\
         \end{array}
       \right],
\end{align}
\begin{align}
\left[%
\begin{array}{c}
{{\bf { h}}_{3,\textrm{R}}^*}[5] \\
{{\bf h}}_{1,\textrm{R}}^*[5]\\
\end{array}%
\right]{\bf { v}}_{2,4}[5]=
\left[
         \begin{array}{c}
           0\\
	{ h}_{1,4}[4]\\
         \end{array}
       \right].
\end{align}
Since we assume that the channel coefficients are drawn from a continuous distribution, we always surely inverse. Therefore, we construct  the relay transmit
beamforming vectors as
\begin{align}
{\bf v}_{3,1}[5]&=\left[%
\begin{array}{c}
{{\bf h}}_{2,\textrm{R}}^*[5]\\
{{\bf { h}}_{4,\textrm{R}}^*}[5] \\
\end{array}%
\right]^{-1}
\left[
     \begin{array}{c}
           0\\
	h_{4,1}[1]\\
         \end{array}
       \right],  \quad
{\bf v}_{4,1}[5]=\left[%
\begin{array}{c}
{{\bf h}}_{1,\textrm{R}}^*[5]\\
{{\bf { h}}_{3,\textrm{R}}^*}[5] \\
\end{array}%
\right]^{-1}
\left[
     \begin{array}{c}
           0\\
	h_{3,1}[2]\\
         \end{array}
       \right], \\
{\bf v}_{3,2}[5]&=\left[%
\begin{array}{c}
{{\bf h}}_{1,\textrm{R}}^*[5]\\
{{\bf { h}}_{4,\textrm{R}}^*}[5] \\
\end{array}%
\right]^{-1}
\left[
         \begin{array}{c}
           0\\
	h_{4,2}[1]\\
         \end{array}
       \right], \quad
{\bf v}_{4,2}[5]=\left[%
\begin{array}{c}
{{\bf h}}_{1,\textrm{R}}^*[5]\\
{{\bf { h}}_{3,\textrm{R}}^*}[5] \\
\end{array}%
\right]^{-1}
\left[
         \begin{array}{c}
           0\\
	h_{3,2}[2]\\
         \end{array}
       \right], \\
{\bf {v}}_{1,3}[5]&=\left[%
\begin{array}{c}
{{\bf { h}}_{4,\textrm{R}}^*}[5] \\
{{\bf h}}_{2,\textrm{R}}^*[5]\\
\end{array}%
\right]^{-1}
\left[
     \begin{array}{c}
           0\\
	{ h}_{2,3}[3]\\
         \end{array}
       \right], \quad
{\bf {v}}_{2,3}[5]=\left[%
\begin{array}{c}
{{\bf { h}}_{4,\textrm{R}}^*}[5] \\
{{\bf h}}_{1,\textrm{R}}^*[5]\\
\end{array}%
\right]^{-1}
\left[
     \begin{array}{c}
           0\\
	{ h}_{1,3}[4]\\
         \end{array}
       \right], \\
{\bf {v}}_{1,4}[5]&=\left[%
\begin{array}{c}
{{\bf { h}}_{3,\textrm{R}}^*}[5] \\
{{\bf h}}_{2,\textrm{R}}^*[5]\\
\end{array}%
\right]^{-1}
\left[
         \begin{array}{c}
           0\\
	{ h}_{2,4}[3]\\
         \end{array}
       \right], \quad
{\bf { v}}_{2,4}[5]=\left[%
\begin{array}{c}
{{\bf { h}}_{3,\textrm{R}}^*}[5] \\
{{\bf h}}_{1,\textrm{R}}^*[5]\\
\end{array}%
\right]^{-1}
\left[
         \begin{array}{c}
           0\\
	{ h}_{1,4}[4]\\
         \end{array}
       \right].
\end{align}
From the relay transmission, the received signals at the users are given by
\begin{eqnarray}
{y}_{1}[5]&=& {{\bf h}_{1,\textrm{R}}^*}[5]\left(\sum_{j=3}^{4}\sum_{i=1}^{2}{\bf v}_{j,i}[5]{s}_{j,i} +  \sum_{j=1}^{2}\sum_{i=3}^{4}{\bf { v}}_{j,i}[5]{{ s}}_{j,i}\right)+z_{1}[5],\nonumber \\
&=&({{\bf h}_{1,\textrm{R}}^*}[5]{\bf v}_{1,3}[5]){s}_{1,3}+({{\bf h}_{1,\textrm{R}}^*}[5]{\bf v}_{1,4}[5]){s}_{1,4}+({{\bf h}_{1,\textrm{R}}^*}[5]{\bf {v}}_{3,1}[5]){{s}}_{3,1}+({{\bf h}_{1,\textrm{R}}^*}[5]{\bf { v}}_{4,1}[5]){{s}}_{4,1}, \nonumber \\
&+&\underbrace{({{\bf h}_{1,\textrm{R}}^*}[5]{\bf {v}}_{2,3}[5]){{ {s}}_{2,3}+({\bf h}_{1,\textrm{R}}^*}[5]{\bf {v}}_{2,4}[5]){{s}}_{2,4}}_{={ h}_{1,3}[4]{s}_{2,3}+{h}_{1,4}[4]{ s}_{2,4}}+z_1[5],  \label{eq:rx1}\end{eqnarray}
\begin{eqnarray}
{y}_{2}[5]&=& {{\bf h}_{2,\textrm{R}}^*}[5]\left(\sum_{j=3}^{4}\sum_{i=1}^{2}{\bf v}_{j,i}[5]{s}_{j,i} +  \sum_{j=1}^{2}\sum_{i=3}^{4}{\bf { v}}_{j,i}[5]{{ s}}_{j,i}\right)+z_{2}[5],\nonumber \\
&=&({{\bf h}_{2,\textrm{R}}^*}[5]{\bf v}_{2,3}[5]){s}_{2,3}+({{\bf h}_{2,\textrm{R}}^*}[5]{\bf v}_{2,4}[5]){s}_{2,4}+\underbrace{({{\bf h}_{2,\textrm{R}}^*}[5]{\bf {v}}_{1,3}[5]){{s}}_{1,3}+({{\bf h}_{2,\textrm{R}}^*}[5]{\bf { v}}_{1,4}[5]){{s}}_{1,4}}_{={ h}_{2,3}[3]{s}_{1,3}+{h}_{2,4}[3]{ s}_{1,4}}, \nonumber \\
&+&({{\bf h}_{2,\textrm{R}}^*}[5]{\bf {v}}_{3,2}[5]){{ {s}}_{3,2}+({\bf h}_{2,\textrm{R}}^*}[5]{\bf {v}}_{4,2}[5]){{s}}_{4,2}+z_2[5],  \label{eq:rx2}\end{eqnarray}
\begin{eqnarray}
{{y}}_{3}[5]&=& {{\bf { h}}_{3,\textrm{R}}^*}[5]\left(\sum_{j=3}^{4}\sum_{i=1}^{2}{\bf v}_{j,i}[5]{s}_{j,i} +  \sum_{j=1}^{2}\sum_{i=3}^{4}{\bf { v}}_{j,i}[5]{{ s}}_{j,i}\right)+{ z}_{3}[5],\nonumber \\
&=&({{\bf { h}}_{3,\textrm{R}}^*}[5]{\bf { v}}_{3,1}[5]){{ s}}_{3,1}+({{\bf h}_{3,\textrm{R}}^*}[5]{\bf { v}}_{3,2}[5]){{ s}}_{3,2}+({{\bf { h}}_{3,\textrm{R}}^*}[5]{\bf {v}}_{1,3}[5]){{s}}_{1,3}+({{\bf { h}}_{3,\textrm{R}}^*}[5]{\bf { v}}_{2,3}[5]){{s}}_{2,3}, \nonumber \\
&+&\underbrace{({{\bf { h}}_{3,\textrm{R}}^*}[5]{\bf {v}}_{4,1}[5]){{ {s}}_{4,1}+({\bf { h}}_{3,\textrm{R}}^*}[5]{\bf {v}}_{4,2}[5]){{s}}_{4,2}}_{={ h}_{3,1}[2]{s}_{4,1}+{h}_{3,2}[2]{ s}_{4,2}}+{ z}_3[5],  \label{eq:rx3}
\end{eqnarray}
\begin{eqnarray}
{{y}}_{4}[5]&=& {{\bf { h}}_{4,\textrm{R}}^*}[5]\left(\sum_{j=3}^{4}\sum_{i=1}^{2}{\bf v}_{j,i}[5]{s}_{j,i} +  \sum_{j=1}^{2}\sum_{i=3}^{4}{\bf { v}}_{j,i}[5]{{ s}}_{j,i}\right)+{ z}_{4}[5],\nonumber \\
&=&({{\bf { h}}_{4,\textrm{R}}^*}[5]{\bf { v}}_{4,1}[5]){{ s}}_{4,1}+({{\bf h}_{4,\textrm{R}}^*}[5]{\bf { v}}_{4,2}[5]){{ s}}_{4,2}+\underbrace{({{\bf { h}}_{4,\textrm{R}}^*}[5]{\bf {v}}_{3,1}[5]){{s}}_{3,1}+({{\bf { h}}_{4,\textrm{R}}^*}[5]{\bf { v}}_{3,2}[5]){{s}}_{3,2}}_{={ h}_{4,1}[1]{s}_{3,1}+{h}_{4,2}[1]{ s}_{3,2}}, \nonumber \\
&+&({{\bf { h}}_{4,\textrm{R}}^*}[5]{\bf {v}}_{2,1}[5]){{ {s}}_{2,1}+({\bf { h}}_{4,\textrm{R}}^*}[5]{\bf {v}}_{2,2}[5]){{s}}_{2,2}+{ z}_4[5]. \label{eq:rx4}
\end{eqnarray}
As shown in (\ref{eq:rx1}), (\ref{eq:rx2}), (\ref{eq:rx3}), and (\ref{eq:rx4}), each user acquires a equation that can be decomposed into three sub-equations, each of which corresponds to desired, self-interference, and aligned interference parts. For instance, for user 1, $({{\bf h}_{1,\textrm{R}}^*}[5]{\bf v}_{1,3}[5]){s}_{1,3}+({{\bf h}_{1,\textrm{R}}^*}[5]{\bf v}_{1,4}[5]){s}_{1,4}$ denotes the desired part as it contains desired information symbols $s_{1,3}$ and $s_{1,4}$.
The sub-equation $({{\bf h}_{1,\textrm{R}}^*}[5]{\bf {v}}_{3,1}[5]){{s}}_{3,1}+({{\bf h}_{1,\textrm{R}}^*}[5]{\bf { v}}_{4,1}[5]){{s}}_{4,1}$ can be interpreted as back propagating self-interference signal from the relay because $s_{3,1}$ and $s_{4,1}$ were transmitted previously by user 1. Last, the sub-equation $({{\bf h}_{1,\textrm{R}}^*}[5]{\bf {v}}_{2,3}[5]){{ {s}}_{2,3}+({\bf h}_{1,\textrm{R}}^*}[5]{\bf {v}}_{2,4}[5]){{s}}_{2,4}$ represents interference signals because ${{s}}_{2,3}$ and ${{s}}_{2,4}$ are intended for user 2. By the proposed precoding, this interference sub-equation has the same shape that was observed at time slot 4 by user 1 in the form of ${ h}_{1,3}[4]{s}_{2,3}+{h}_{1,4}[4]{ s}_{2,4}$.

\subsubsection{Decoding}
Let us explain the decoding procedure for user 1. First, user 1 eliminate the back propagating self-interference signals $M_1[5]=({{\bf h}_{1,\textrm{R}}^*}[5]{\bf {v}}_{3,1}[5]){{s}}_{3,1}+({{\bf h}_{1,\textrm{R}}^*}[5]{\bf { v}}_{4,1}[5]){{s}}_{4,1}$ from $y_1[5]$ by using knowledge of the effective channel ${{\bf h}_{1,\textrm{R}}^*}[5]{\bf v}_{3,1}[5]$ and ${{\bf h}_{1,\textrm{R}}^*}[5]{\bf v}_{4,1}[5]$ and the transmitted data symbols $s_{3,1}$ and $s_{4,1}$. Second, user 1 removes the effect of interference $({{\bf h}_{1,\textrm{R}}^*}[5]{\bf {v}}_{2,3}[5]){{ {s}}_{2,3}+({\bf h}_{1,\textrm{R}}^*}[5]{\bf {v}}_{2,4}[5]){{s}}_{2,4}$ by using the fact that $({{\bf h}_{1,\textrm{R}}^*}[5]{\bf {v}}_{2,3}[5]){{ {s}}_{2,3}+({\bf h}_{1,\textrm{R}}^*}[5]{\bf {v}}_{2,4}[5]){{s}}_{2,4}=L_1[4]$.
After canceling the known interference, the concatenated input-output relationship seen by user 1 is
\begin{eqnarray}
\left[%
\begin{array}{c}
  {y}_1[3] \\
  {y}_1[5]-{L}_1[4]-M_1[5] \\
\end{array}%
\right]=\underbrace{\left[%
\begin{array}{cc}
  h_{1,3}[3] & h_{1,4}[3] \\
 {{\bf h}_{1,\textrm{R}}^*}[5]{\bf v}_{1,3}[5] & {{\bf h}_{1,\textrm{R}}^*}[5]{\bf v}_{1,4}[5]\\
\end{array}%
\right]}_{{\bf \tilde{H}}_{1}}\left[%
\begin{array}{c}
  s_{1,3} \\
  s_{1,4} \\
\end{array}%
\right]+\left[%
\begin{array}{c}
  z_1[3] \\
  z_1[5]-z_1[4] \\
\end{array}%
\right].
\end{eqnarray}
Since beamforming vectors, ${\bf v}_{1,3}[5]$ and ${\bf v}_{1,4}[5]$, were constructed independently from the direct channel ${ h}_{1,3}[3]$ and $h_{1,4}[3]$, then, $\textrm{rank}\left({\bf \tilde{H}}_{1}\right)=2$. As a result, user 1 decodes two desired symbols $s_{1,3}$ and $s_{1,4}$ based on five channel uses. Similarly, the other users decode two desired information symbols by using the same method. Consequently, a total eight data symbols have been delivered in five channel uses in the network, implying that a total $d_{\textrm{sum}}=\frac{8}{5}$ is achieved.

\section{Extension to Multiple Relays with a Single Antenna}
So far, the scenario considered in this paper used a single relay with $N$ antennas. This relay can be viewed as $N$ relays with a single antenna where the relays fully cooperate by sharing both data and CSI. In this section, we consider multiple distributed relays which of each has a single antenna. We assume that the relays have full CSI but do not share data. By providing two examples, we show the multiple relays with a single antenna can increase the DoF gain for multi-way interference networks.

\begin{figure}
\centering
\includegraphics[width=4in]{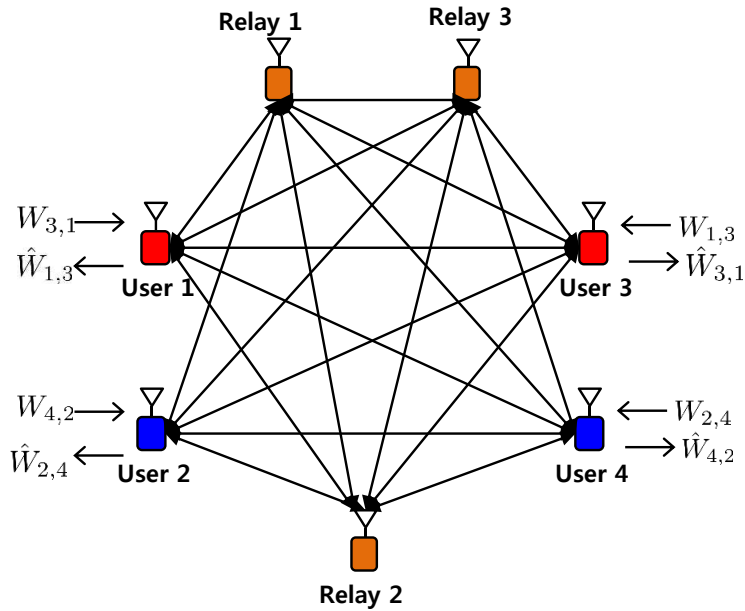}
\caption{The two-pair two-way interference channel with three disturbed relays employing a single antenna.} \label{fig:7}
\end{figure}

\subsection{Example 3: Two-Pair Two-Way Intererference Channel with Three Distributed Relays }
In this example, we consider a two-pair two-way interference channel with three distributed relays as illustrated in Fig. \ref{fig:7}. Note that this network is similar to Example 1 in Section III,  except that there are single-antenna relays instead of one two-antenna relay. In this example, we prove the following theorem.

 \begin{theorem} \label{Theorem3}
For the two-pair two-way interference channel with three distributed relays employing a single antenna, a total $\frac{4}{3}$ of DoF is achievable.
\end{theorem}

\textit{Proof}: The achievability is shown by the proposed space-time interference neutralization.

\subsubsection{Phase 1 (Forward IC Transmission)} In the first time slot, user 1 and user 2 send signals $x_1[1]=s_{3,1}$ and $x_2[1]=s_{4,2}$. Ignoring noise, the received signals at user 3, user 4, and the relays are given by
\begin{eqnarray}
{ y}_{3}[1] &=&{h}_{3,1}[1]s_{3,1}+  {h}_{3,2}[1]s_{4,2}, \label{eq:dis2} \\
{ y}_{4}[1] &=& {h}_{4,1}[1]s_{3,1}+  {h}_{4,2}[1]s_{4,2}, \label{eq:dis1}\\
{ y}^n_{\textrm{R}}[1]&=&{ h}^n_{\textrm{R},1}[1]s_{3,1} +{ h}^n_{\textrm{R},2}[1] s_{4,2},  \quad n\in\mathcal{R}=\{1,2,3\} \label{eq:dis3}
\end{eqnarray}
where ${ y}^n_{\textrm{R}}[1]$ and ${h}^n_{\textrm{R},i}[1]$ denote the received signal at relay $n\in \mathcal{R}$ and the channel value from user $i$ to relay $n$ at time slot 1. In contrast to Example 1, the relays do not decode the transmitted data symbols $s_{3,1}$ and $s_{4,2}$ because they have only a single antenna.

\subsubsection{Phase 2 (Backward IC Transmission)} In the second time slot, user $3$ and user $4$ send signals ${x}_{3}[2]={ s}_{1,3}$ and ${ x}_{4}[2]={s}_{{2,4}}$ over the backward interference channel. The received signals at user 1, user 2, and the three relays are given by
\begin{eqnarray}
y_1[2]&=&{ h}_{1,3}[2]{ s}_{1,3}+{ h}_{1,4}[2]{ s}_{{2,4}} \nonumber \\
y_2[2]&=&{ h}_{2,3}[2]{ s}_{{1,3}}+{ h}_{2,4}[2]{ s}_{{2,4}}\nonumber \\
{ y}^n_{\textrm{R}}[2]&=&{ { h}}^n_{\textrm{R},3}[2]{ s}_{1,3} +{ { h}}^n_{\textrm{R},4}[2] { s}_{{2,4}}.
\end{eqnarray}

\subsubsection{Phase 3 (The Relay Broadcast) } In the third time slot, the three relays cooperatively send the received signals for the previous time slots. The signal transmitted by relay $n\in \mathcal{R}$ is
 \begin{align}
{ x}^n_R[3]&={ v}^n[1]y_R^n[1] +{v}^n[2]y_R^n[2] \\
&={ v}^n[1]({ h}^n_{\textrm{R},1}[1]s_{3,1} +{ h}^n_{\textrm{R},2}[1] s_{4,2}) +{v}^n[2]({ { h}}^n_{\textrm{R},3}[2]{ s}_{1,3} +{ { h}}^n_{\textrm{R},4}[2] { s}_{{2,4}}),
 \end{align}
where $v^n[t]$ denotes a relay precoding coefficient used for the received signal $y_R^n[t]$ for $t\in \{1,2\}$.

Let ${\bf g}_{\ell,\textrm{R},i}^*=[{ h}^{1}_{\ell,\textrm{R}}[3]{ h}^1_{\textrm{R},i}[1],~{ h}^{2}_{\ell,\textrm{R}}[3]{ h}^2_{\textrm{R},i}[1],~{ h}^{3}_{\ell,\textrm{R}}[3]{ h}^3_{\textrm{R},i}[1] ]$ for $i\in\{1,2\}$ and ${\bf g}_{\ell,\textrm{R},j}^*=[{ h}^{1}_{\ell,\textrm{R}}[3]{ h}^1_{\textrm{R},j}[2],~{ h}^{2}_{\ell,\textrm{R}}[3]{ h}^2_{\textrm{R},j}[2],~{ h}^{3}_{\ell,\textrm{R}}[3]{ h}^3_{\textrm{R},j}[2] ]$ for $j\in\{3,4\}$ denote effective channel vector from user $i$ ($j$) to user $\ell \in\{1,2,3,4\}$ via the three relays and ${\bf v}[t]=[v^{1}[t],~ v^{2}[t],~ v^{3}[t]]^*$ for $t\in\{1,2\}$ represents precoding vector used by the three relays in time slot $t$. The received signal at user $\ell\in \{1,2,3,4\}$ in time slot 3 is given by
\begin{eqnarray}
{y}_{\ell}[3]&=& \sum_{n=1}^3{ h}^{n}_{\ell,\textrm{R}}[3]{x}^n_{\textrm{R}}[3]  \\
&=& \sum_{n=1}^3{ h}^{n}_{\ell,\textrm{R}}[3]\left\{{ v}^n[1]({ h}^n_{\textrm{R},1}[1]s_{3,1} +{ h}^n_{\textrm{R},2}[1] s_{4,2}) +{v}^n[2]({ { h}}^n_{\textrm{R},3}[2]{ s}_{1,3} +{ { h}}^n_{\textrm{R},4}[2] { s}_{{2,4}})\right\} \\
&=&{\bf g}_{\ell,\textrm{R},1}^*{\bf v}[1]s_{3,1}+{\bf g}_{\ell,\textrm{R},2}^*{\bf v}[1]s_{4,2}+{\bf g}_{\ell,\textrm{R},3}^*{\bf v}[2]s_{1,3}+{\bf g}_{\ell,\textrm{R},4}^*{\bf v}[2]s_{2,4},
\end{eqnarray}
The role of the three relays in this example is to control the information flow so that each user does not receive unknown interference signals. For instance, user 1 and user 2 do not have knowledge of interference symbol $s_{4,2}$ and $s_{3,1}$. Similarly, user 3 and user 4 do not know interference symbol $s_{2,4}$ and $s_{1,3}$, respectively. Thus, the following interference neutralization conditions are required
\begin{align}
\left[%
\begin{array}{c}{\bf g}_{1,\textrm{R},2}^* \\
 {\bf g}_{2,\textrm{R},1}^* \\
\end{array}%
\right]{\bf v}[1]={\bf 0}_{2 \times 1}\quad \textrm{and} \quad \left[%
\begin{array}{c}{\bf g}_{3,\textrm{R},4}^* \\
 {\bf g}_{4,\textrm{R},3}^* \\
\end{array}%
\right]{\bf v}[2]={\bf 0}_{2 \times 1}. \label{IN}
\end{align}
Note that the precoding solutions for ${\bf v}[1]$ and ${\bf v}[2]$ exist always in this case from the existence of null space of the matrices $ \left[%
\begin{array}{c}{\bf g}_{1,\textrm{R},2}^* \\
 {\bf g}_{2,\textrm{R},1}^* \\
\end{array}%
\right] \in \mathbb{C}^{2\times 3}$ and $ \left[%
\begin{array}{c}{\bf g}_{3,\textrm{R},4}^* \\
 {\bf g}_{4,\textrm{R},3}^* \\
\end{array}%
\right] \in \mathbb{C}^{2\times 3}$.

\subsubsection{Decoding} To explain the decoding, we consider user 1. From the interference neutralization conditions in (\ref{IN}), the received signal is
\begin{eqnarray}
{y}_{1}[3]&=&{\bf g}_{1,\textrm{R},1}^*{\bf v}[1]s_{3,1}+{\bf g}_{1,\textrm{R},3}^*{\bf v}[2]s_{1,3}+{\bf g}_{1,\textrm{R},4}^*{\bf v}[2]s_{2,4}.
\end{eqnarray}
We first subtract off contribution from the known signals to form ${y}_{1}[3]-{\bf g}_{1,\textrm{R},1}^*{\bf v}[1]s_{3,1}$. Then, the concatenated input-output relationship is given by
\begin{eqnarray}
\left[%
\begin{array}{c}
  {y}_{1}[2]\\
 {y}_{1}[3]-{\bf g}_{1,\textrm{R},1}^*{\bf v}[1]s_{3,1}\\
\end{array}%
\right]=\underbrace{\left[%
\begin{array}{cc}
 {h}_{1,3}[2] & { h}_{1,4}[2] \\
{\bf g}_{1,\textrm{R},3}^*{\bf v}[2]& {\bf g}_{1,\textrm{R},4}^*{\bf v}[2]\\
\end{array}%
\right]}_{{\bf \tilde{G}}_{1}}\left[%
\begin{array}{c}
  { s}_{1,3} \\
 { s}_{2,4} \\
\end{array}%
\right] .
\end{eqnarray}
Since the effective channel matrix ${\bf \tilde{G}}_{1}$ has full rank almost surely, user 1 decodes the desired symbol ${ s}_{1,3}$ by applying a ZF decoder. By symmetry, the other users operate in a similar fashion, which implies that a total $\frac{4}{3}$ of DoF is achievable.

\textit{Remark 7 (Comparision with results in \cite{Rankov})} Interference neutralization method was introduced in \cite{Rankov} for the $K \times L \times K$ layered two-hop interference network. To realize interference neutralization for bi-directional information exchange between $K$-paris (2$K$ users), $L\geq 2K(K-1)+1$ relays with a single antenna are needed. For example, when $K=2$, the minimum required relays are $L=5$ to achieve $2$ DoF for the layered two-hop half-duplex interference network. When the feasibility condition of interference neutralization does not hold, i.e., $L<5$, there are no results so far that the relays increase the achievable DoF beyond one. Our result, however, shows that with fewer than five relays, it is possible to increase the achievable DoF beyond one by using what we call \textit{space-time interference neutralization}.


\subsection{Example 4: Three-User Fully-Connected Y Channel with Three Single Antenna Relays}
\begin{figure}
\centering
\includegraphics[width=4in]{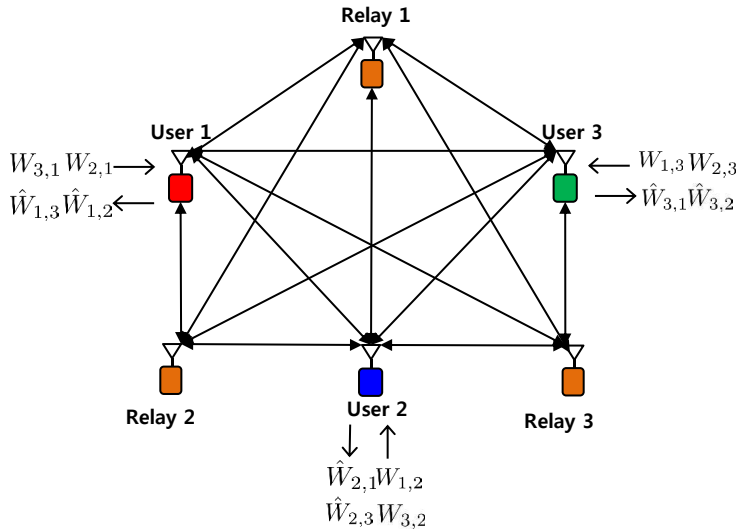}
\caption{3-user fully-connected Y channel with three distributed relays employing a single antenna.} \label{fig:ychannel_dis}
\end{figure}

Now, we consider a fully-connected 3-user Y channel with three distributed single-antenna relays as depicted in Fig. \ref{fig:ychannel_dis}. By using space-time interference neutralization explained in Example 3, we show the following theorem.

 \begin{theorem} \label{Theorem4}
For the 3-user fully-connected Y channel with three distributed single-antenna relays, the optimal $\frac{3}{2}$ of DoF is achievable.
\end{theorem}
 \textit{Proof}: Since the converse argument is direct from Theorem 1, achievability is shown by two-phase communication protocol. The  first phase comprises three time slots, i.e., $T_1=\{1,2,3\}$. For $k\in T_1$ time slot, all users excepting user $k$ transmit signals intended for user $k$. Thus, the received signals at user $k$ and relay $n\in \{1,2,3\}$ are given by
\begin{align}
{y}_{k}[k] &= \sum_{\ell=1,\ell\neq k}^{3}{h}_{k,\ell}[k]s_{k,\ell} + {z}_{k}[k], \\
{ y}^n_{\textrm{R}}[k] &= \sum_{\ell=1,\ell\neq k}^{3}{ h}^n_{\textrm{R},\ell}[k]s_{k,\ell} + { z}^n_{\textrm{R}}[k].
\end{align}
During phase one, each user $k$ has one desired equation and each relay has three equations. In the second phase, one time slot is used for the relay transmission, $T_2=\{4\}$. During time slot 4, the three relays cooperatively send signals to the users based on what they obtained during the first phase. The transmitted signal at the $n$-th relay is given by
\begin{align}
{ x}^n_R[4]= \sum_{k=1}^{3}{v}^n[k]y^n_R[k],
\end{align}
where ${v}^n[k]$ denotes the
precoding coefficient used at the $n$-th relay for the $k$-th time slot observation $y^n_R[k]$. When the three relays send at time slot 4, the received signal at user $j$ is given by
\begin{align}
y_{j}[4] &=\sum_{n=1}^{3}{ h}^n_{j,\textrm{R}}[4]{ x}^n_R[4] \nonumber \\
&=\sum_{n=1}^{3}{ h}^n_{j,\textrm{R}}[4]\sum_{k=1}^{3}v^n[k]y^n_R[k] \nonumber \\
&=\sum_{n=1}^{3}{ h}^n_{j,\textrm{R}}[4]\sum_{k=1}^{3}v^n[k]\left( \sum_{\ell=1,\ell\neq k}^{3}{ h}^n_{\textrm{R},\ell}[k]s_{k,\ell}\right) \\
&={\bf h}_{j,\textrm{R}}^*[4]{\bf V}^R\left[\!\!%
\begin{array}{cccccc}
  {\bf H}_{\textrm{R},\mathcal{U}_1^{\textrm{c}}}[1] & {\bf H}_{\textrm{R},\mathcal{U}_2^{\textrm{c}}}[2]  & {\bf H}_{\textrm{R},\mathcal{U}_3^{\textrm{c}}}[3]
\end{array}\!\!%
\right]\left[\!\!%
\begin{array}{c}
  {\bf s}_{1,\mathcal{U}_1^{\textrm{c}}} \\
 {\bf s}_{2,\mathcal{U}_2^{\textrm{c}}} \\
 {\bf s}_{3,\mathcal{U}_3^{\textrm{c}}}
\end{array}\!\!%
\right],  \label{eq:Y_1}
\end{align}
where ${\bf h}_{j,\textrm{R}}^*[4]=[h_{j,\textrm{R}}^1[4],h_{j,\textrm{R}}^1[4],h_{j,\textrm{R}}^3[4]] \in \mathbb{C}^{1\times 3}$ denotes the channel vector from the three relays to user $j\in\{1,2,3\}$ at time slot $4$; ${\bf V}^R \in \mathbb{C}^{3\times 3}$ denotes a space-time relay network coding  matrix at time slot $4$ whose $(k,n)$ element is defined as ${\bf V}^R(n,k)=[v^n[k]]$; ${\bf H}_{\textrm{R},\mathcal{U}_k^{\textrm{c}}}[k] \in \mathbb{C}^{3 \times 2}$ denotes the effective channel matrix from user group $\mathcal{U}_{k}^{\textrm{c}}$ to the three relays at time slot $k\in T_1$; ${\bf s}_{j,\mathcal{U}_j^{\textrm{c}}}$ represents the desired symbol vector at user $j$, which comes from user group $\mathcal{U}_j^{\textrm{c}}$. Note that user $j$ receives a linear combination of six symbols from the relays. Let us decompose these six symbols into three terms: two desired symbols ${\bf s}_{j,\mathcal{U}_j^{\textrm{c}}}$, two self-interference symbols ${\bf s}_{\mathcal{U}_j^{\textrm{c}},j}$, and two inter-user interference symbols ${\bf s}_{I^{\textrm{c}}_j}=[s_{k,\ell}] $ where $k\neq j$ or $\ell \neq j$. Then, let define a permutation matrix ${\bf P}_j\in \mathbb{Z}^{6 \times 6}$ that changes the order of transmitted symbols such that
\begin{align}
\left[\!\!%
\begin{array}{c}
  {\bf s}_{1,\mathcal{U}_1^{\textrm{c}}} \\
 {\bf s}_{2,\mathcal{U}_2^{\textrm{c}}} \\
 {\bf s}_{K,\mathcal{U}_3^{\textrm{c}}}
\end{array}\!\!%
\right]&= {\bf P}_j \left[\!\!%
\begin{array}{c}
{\bf s}_{j,\mathcal{U}_j^{\textrm{c}}} \\
{\bf s}_{\mathcal{U}_j^{\textrm{c}},j} \\
{\bf s}_{I^{\textrm{c}}_j} \\
\end{array}\!\!%
\right].
\end{align}
Using the permutation matrix, we can rewrite the equation in (\ref{eq:Y_1}) as
\begin{align}
y_{j}[4] &={\bf h}_{j,\textrm{R}}^*[4]{\bf V}^R\left[\!\!%
\begin{array}{cccccc}
  {\bf H}_{\textrm{R},\mathcal{U}_1^{\textrm{c}}}[1] & {\bf H}_{\textrm{R},\mathcal{U}_2^{\textrm{c}}}[2] &  {\bf H}_{\textrm{R},\mathcal{U}_3^{\textrm{c}}}[3]
\end{array}\!\!%
\right]{\bf P}_j \left[\!\!%
\begin{array}{c}
{\bf s}_{j,\mathcal{U}_j^{\textrm{c}}} \\
{\bf s}_{\mathcal{U}_j^{\textrm{c}},j} \\
{\bf s}_{I^{\textrm{c}}_j} \\
\end{array}\!\!%
\right] \\
&={\bf h}_{j,\textrm{R}}^*[4]{\bf V}^R{\bf A}_j{\bf s}_{j,\mathcal{U}_j^{\textrm{c}}} + {\bf h}_{j,\textrm{R}}^*[t]{\bf V}^R{\bf B}_j{\bf s}_{\mathcal{U}_j^{\textrm{c}},j} +{\bf h}_{j,\textrm{R}}^*[t]{\bf V}^R{\bf C}_j{\bf s}_{I^{\textrm{c}}_j} , \label{eq:Y_2}
\end{align}
where the effective channel matrices  ${\bf A}_j\in \mathbb{C}^{3\times 2}$, ${\bf B}_j\in \mathbb{C}^{3 \times 2}$, and ${\bf C}_j\in \mathbb{C}^{3 \times 2}$ are defined as
\begin{align}
\left[\!\!%
\begin{array}{ccc}
  {\bf A}_j & {\bf B}_j & {\bf C}_j
\end{array}\!\!%
\right]&=\left[\!\!%
\begin{array}{cccccc}
  {\bf H}_{\textrm{R},\mathcal{U}_1^{\textrm{c}}}[1] & {\bf H}_{\textrm{R},\mathcal{U}_2^{\textrm{c}}}[2] &  {\bf H}_{\textrm{R},\mathcal{U}_3^{\textrm{c}}}[3]
\end{array}\!\!%
\right]{\bf P}_j .
\end{align}
To eliminate interference signal ${\bf s}_{I^{\textrm{c}}_j}$ for user $j$, the relays cooperatively design space-time relay network coding matrix ${\bf V}^R$ so that the following interference neutralization condition is satisfied,
\begin{align}
{\bf h}_{j,\textrm{R}}^*[4]{\bf V}^R{\bf C}_j={\bf 0}_{1\times 2}, \quad \textrm{for} \quad j \in \mathcal{U}.\label{eq:IN_STIN1}
\end{align}
To solve the matrix equations in (\ref{eq:IN_STIN1}) for all $K$ users, we convert them into vector forms by exploiting Kronecker product operation property,
$\textrm{vec}({\bf AXB})=({\bf B}^T\otimes{\bf
A})\textrm{vec}({\bf X})$. The combined vector form of
interference neutralization in (\ref{eq:IN_STIN1}) is given by
\begin{align}
\underbrace{\left[\!\!%
\begin{array}{c}
{\bf C}_1^T\otimes{\bf h}_{1,\textrm{R}}^*[4] \\
{\bf C}_2^T\otimes{\bf h}_{2,\textrm{R}}^*[4]\\
{\bf C}_3^T\otimes{\bf h}_{3,\textrm{R}}^*[4] \\
\end{array}\!\!%
\right]}_{{\bf \bar{F}} \in \mathbb{C}^{8\times 9}}\underbrace{{\bf \bar{v}}^R}_{9\times 1} ={\bf 0}_{8\times 1}.
\end{align}
where ${\bf \bar{v}}^R=\textrm{vec}({\bf V}^R)$ is the vector representation of relay beamforming matrix
${\bf V}^R$ by stacking the column vectors of it.
Because the elements of the channels are drawn from
a continuous random variable and the size of the unified system matrix
${\bf \bar{F}}$ is $8 \times 9$, ${\bf \bar{F}}$ has a null subspace almost surely. Therefore, the relay
beamforming vector eliminating all interference signals on the network are obtained as
\begin{eqnarray}
{\bf \bar{v}}^R \in {\textrm{null}}({\bf \bar{F}}).
\end{eqnarray}
By reshaping the vector solution ${\bf \bar{v}}^R$ into a matrix, we obtain the network-wise space-time relay precoding matrix ${\bf V}^R$.

Last, let us consider the decoding procedure at user $1$. Recall that user $1$ received an equation consisting of two desired symbols ${\bf s}_{1,\mathcal{U}_1^{\textrm{c}}}=[s_{1,2} ~s_{1,3}]^T$ in time slot 1. In addition, in time slot 4, user 1 received a signal from the relay in the form of $y_1[4]={\bf h}_{1,\textrm{R}}^*[4]{\bf V}^R{\bf A}_1{\bf s}_{1,\mathcal{U}_1^{\textrm{c}}} + {\bf h}_{1,\textrm{R}}^*[4]{\bf V}^R{\bf B}_1{\bf s}_{\mathcal{U}_1^{\textrm{c}},1} $.  Since user 1 has knowledge of ${\bf s}_{\mathcal{U}_1^{\textrm{c}},1} $, it cancels known interference symbols from $y_1[4]$. Thus, the input-out relationship is given by
\begin{align}
\left[%
\begin{array}{c}
  {y}_{1}[1]\\
 {y}_{1}[4]- {\bf h}_{1,\textrm{R}}^*[4]{\bf V}^R{\bf B}_1{\bf s}_{\mathcal{U}_1^{\textrm{c}},1}\\
\end{array}%
\right]=\underbrace{\left[%
\begin{array}{c}
 {\bf h}_{1,\mathcal{U}_1^{\textrm{c}}}[1] \\
{\bf h}_{1,\textrm{R}}^*[4]{\bf V}^R{\bf A}_1\\
\end{array}%
\right]}_{{\bf \tilde{H}}_{1}}{\bf s}_{1,\mathcal{U}_1^{\textrm{c}}},
\end{align}
where ${\bf h}_{1,\mathcal{U}_1^{\textrm{c}}}[1]=[{ h}_{1,2}[1],~h_{1,3}[1]]$.
Since the effective channel matrix ${\bf \tilde{H}}_{1}$ has full rank two, user 1 decodes two desired symbols. By the symmetry, user 2 and 3 obtain their desired information symbols as well. As a result, it is possible to achieve $\frac{3}{2}$ DoF. This completes the proof.

\section{Conclusion}
In this paper, we studied a fully-connected interference network with relay nodes. By considering the multi-way information flows, we characterized the sum-DoF for the fully-connected Y channel by yielding converse based on the cut-set theorem and achievability based on the proposed multi-phase transmission scheme. Further, we derived DoF inner bounds for different message setups and multiple relays with a single antenna case in the four-user fully-connected interference network. From the derived DoF results, we verified an intuition that relays are useful in increasing the DoF of multi-user interference network when the multi-way information flows are considered, even if the relays and users operate in half-duplex. This DoF increase is due to two gains: the caching gain that inherently given by multi-way communications and the interference shaping gain by the space-time relay transmission technique that controls information flow in the network.


\appendices
\section{Proof of Lemma 1}
We start by arguing that allowing the
cooperation of among users does not reduce the DoF. Using this fact, the cooperation between all users excepting user 1 is considered. In this case, the cooperating
user group consisted of user $k \in \mathcal{U}_{1}^{\textrm{c}}$ where $\mathcal{U}_{1}^{\textrm{c}}=\mathcal{U}/\{1\}$ intends to send messages $W_{1,k}$ $k \in \mathcal{U}_{1}^{\textrm{c}}$ to user 1. User 1 also intends to send the messages $W_{k,1}$ $k\in\mathcal{U}_{1}^{\textrm{c}}$ to the user group. This user cooperation changes the $K$-user fully-connected Y channel into the two-way relay channel where user group $\mathcal{U}_{1}^{\textrm{c}}$ has $K-1$ antennas but user 1 has a single antenna. In this half-duplex non-separated two-way relay channel scenario, we need to show that $\sum_{k=2}^{K}d_{k,1} + \sum_{\ell=2}^{K}d_{1,\ell} \leq 1$.

In general, for the half-duplex non-separated two-way relay channel as illustrated Fig. \ref{upper}, six different network states exist. Let $\lambda_i$, $i \in \{1,2,\ldots, 6 \}$ denote the fraction of transmission time used for network state $i$. For each state, the six- phase transmissions can occur. In the first phase $n\in\lambda_1$, the user group transmits, while both the relay and user 1 listen. Alternatively, in the second phase $n\in\lambda_2$, user 1 sends the signal, while the relay and the user group listen. In the third phase $n\in\lambda_3$, the user group and user 1 transmit the signal at the same time to the relay. For the fourth phase $n\in\lambda_4$, the relay helps the transmission of the user group. In the fifth phase $n\in\lambda_5$, it also assists the transmission of user 1. Lastly, the relay broadcasts the signal to both the user group and user 1 for the sixth phase $n\in\lambda_6$. For the half-duplex relay network with six states, we apply the cut-set theorem \cite{cut-set} to obtain the upper bounds of information flow from the user group to user 1 $R_{1,\mathcal{U}_{1}^{\textrm{c}}} $ and from user 1 to the user group $R_{\mathcal{U}_{1}^{\textrm{c}},1} $. Let us first consider the upper bound of the rate $R_{1,\mathcal{U}_{1}^{\textrm{c}}} $, which is
\begin{eqnarray}
R_{1,\mathcal{U}_{1}^{\textrm{c}}} \leq \min \{ R_{1,\mathcal{U}_{1}^{\textrm{c}}}^S, R_{1,\mathcal{U}_{1}^{\textrm{c}}}^D \},
\end{eqnarray}
where $R_{B,A}^S$ and $R_{B,A} ^D$ denote information transfer rates from $A$ to $B$ across the cut around the source and destination nodes, which are defined as
\begin{align}
R_{1,\mathcal{U}_{1}^{\textrm{c}}}^S &= \lambda_1I({\bf x}_{\mathcal{U}_{1}^{\textrm{c}}};{\bf y}_R, {\bf y}_1,i=1) + \lambda_3I({\bf x}_{\mathcal{U}_{1}^{\textrm{c}}};{\bf y}_R |{\bf x}_1,i=3)+ \lambda_5I({\bf x}_{\mathcal{U}_{1}^{\textrm{c}}};{\bf y}_1| {\bf x}_R,i=5), \label{eq:up1} \\
R_{1,\mathcal{U}_{1}^{\textrm{c}}}^D &= \lambda_1I({\bf x}_{\mathcal{U}_{1}^{\textrm{c}}}; {\bf y}_1,i=1) + \lambda_4I({\bf x}_{\textrm{R}};{\bf y}_1,i=4)+ \lambda_5I({\bf x}_{\mathcal{U}_{1}^{\textrm{c}}}, {\bf x}_R;{\bf y}_1,i=5),\label{eq:up2}
\end{align}
where ${\bf x}_{\mathcal{U}_{1}^{\textrm{c}}}=\{x_2[n], \ldots, x_{K}[n]\}$, ${\bf x}_{1}=\{x_1[n]\}$, ${\bf x}_{\textrm{R}}=\{y_1[n]\}$, ${\bf y}_{1}=\{y_1[n]\}$, and ${\bf y}_{\textrm{R}}=\{ {\bf y}_R[n]\}$ for $n\in \lambda_i$. Also, $I({\bf a};{\bf b})$ and $I({\bf a};{\bf b}|{\bf c})$ denote mutual information between two random vectors ${\bf a}$ and ${\bf b}$ and mutual information conditioned on random vector ${\bf c}$. Similarly, we have the upper bound of the rate $R_{\mathcal{U}_{1}^{\textrm{c}},1}$ as \begin{eqnarray}
R_{\mathcal{U}_{1}^{\textrm{c}},1} \leq \min \{R_{\mathcal{U}_{1}^{\textrm{c}},1} ^S,\textrm{R}_{\mathcal{U}_{1}^{\textrm{c}},1}^D\},
\end{eqnarray}
where
\begin{align}
R_{\mathcal{U}_{1}^{\textrm{c}},1}^S &= \lambda_2I({\bf x}_{1};{\bf y}_R, {\bf y}_{\mathcal{U}_{1}^{\textrm{c}}},i=2) + \lambda_3I({\bf x}_{1};{\bf y}_R|{\bf x}_{\mathcal{U}_{1}^{\textrm{c}}},i=3)+ \lambda_6I({\bf x}_{1};{\bf y}_{\mathcal{U}_{1}^{\textrm{c}}}| {\bf x}_R,i=6), \label{eq:up3} \\
R_{\mathcal{U}_{1}^{\textrm{c}},1}^D &= \lambda_2I({\bf x}_{1}; {\bf y}_{\mathcal{U}_{1}^{\textrm{c}}},i=2) + \lambda_4I({\bf x}_{\textrm{R}};{\bf y}_{\mathcal{U}_{1}^{\textrm{c}}},i=4)+ \lambda_6I({\bf x}_{1}, {\bf x}_R;{\bf y}_{\mathcal{U}_{1}^{\textrm{c}}},i=6). \label{eq:up4}
\end{align}
From each mutual information expression term in (\ref{eq:up1}), (\ref{eq:up2}), (\ref{eq:up3}), and (\ref{eq:up4}), we have the DoF upper bounds of the rates $R_{1,\mathcal{U}_{1}^{\textrm{c}}}^S$, $R_{1,\mathcal{U}_{1}^{\textrm{c}}}^D$,  $R_{\mathcal{U}_{1}^{\textrm{c}},1}^S$, and $R_{\mathcal{U}_{1}^{\textrm{c}},1}^D$ as
\begin{align}
\lim_{P\rightarrow \infty  } \frac{R_{1,\mathcal{U}_{1}^{\textrm{c}}}^S}{\log P} &\leq \lambda_1 \min\{K-1,N+1\}+\lambda_3 \min\{K-1,N\}+\lambda_5\min\{K-1,1\}\\
&=  \lambda_1 (K-1)+\lambda_3 (K-1)+\lambda_5, \\
\lim_{P\rightarrow \infty  } \frac{R_{1,\mathcal{U}_{1}^{\textrm{c}}}^D}{\log P} &\leq \lambda_1 \min\{K-1,1\}+\lambda_4 \min\{N,1\}+\lambda_5\min\{K-1+N,1\}\\
&=  \lambda_1+\lambda_4+\lambda_5, \\
\lim_{P\rightarrow \infty  } \frac{R_{\mathcal{U}_{1}^{\textrm{c}},1}^S}{\log P} &\leq \lambda_2 \min\{1,K-1+N\}+\lambda_3 \min\{1,N\}+\lambda_6\min\{1,K-1\}\\
&=  \lambda_2 +\lambda_3 +\lambda_6, \\
\lim_{P\rightarrow \infty  } \frac{R_{\mathcal{U}_{1}^{\textrm{c}},1}^D}{\log P} &\leq \lambda_2 \min\{1,K-1\}+\lambda_4 \min\{N,K-1\}+\lambda_6\min\{N+1,K-1\}\\
&=  \lambda_2+\lambda_4(K-1)+\lambda_6(K-1),
\end{align}
where the equalities follows from $N < K$.
Using above results, the maximum DoF for
information transfer from user group $\mathcal{U}_{1}^{\textrm{c}}$ to user 1 is
\begin{eqnarray}
\sum_{k=2}^{K}d_{1,k} &= & \min\left\{\lim_{P\rightarrow \infty  } \frac{R_{1,\mathcal{U}_{1}^{\textrm{c}}}^S}{\log P} ,\lim_{P\rightarrow \infty  } \frac{R_{1,\mathcal{U}_{1}^{\textrm{c}}}^D}{\log P} \right\}, \\
&\leq& \min\{(K-1)(\lambda_1+\lambda_3)+\lambda_5, ~~ \lambda_1+\lambda_4+\lambda_5\}.\label{eq:upper1}
\end{eqnarray}
Similarly, the maximum DoF for
information transfer from user 1 to user group $\mathcal{U}_{1}^{\textrm{c}}$ is
\begin{eqnarray}
\sum_{k=2}^{K}d_{k,1} &= & \min\left\{\lim_{P\rightarrow \infty  } \frac{R_{\mathcal{U}_{1}^{\textrm{c}},1}^S}{\log P} ,\lim_{P\rightarrow \infty  } \frac{R_{\mathcal{U}_{1}^{\textrm{c}},1}^D}{\log P} \right\},
\\ &\leq& \min\{\lambda_2+\lambda_3+\lambda_6, ~~ \lambda_2+(K-1)(\lambda_4+\lambda_6)\}. \label{eq:upper2}
\end{eqnarray}
Using (\ref{eq:upper1}) and (\ref{eq:upper2}), an upper bound of the sum-DoF is obtained by solving the following linear program,
\begin{align}
\max_{\lambda_1,\ldots,\lambda_6} &\min\{(K\!-\!1)(\lambda_1\!+\!\lambda_3)\!+\!\lambda_5,~ \lambda_1\!+\!\lambda_4\!+\!\lambda_5\} + \min\{\lambda_2\!+\!\lambda_3+\lambda_6,~ \lambda_2\!+\!(K\!-\!1)(\lambda_4\!+\!\lambda_6) \} \nonumber \\
& \textrm{subject to} \quad  \sum_{i=1}^{6}\lambda_i=1, \quad \lambda_i \geq 0. \label{eq:obj}
\end{align}
Although the optimal value of this linear programing problem can be obtained by using optimization techniques accurately, we can simply find the upper bound of the optimal value by using the fact that $\min \{\alpha,\beta\} + \min \{\delta,\gamma\} \leq \min \{\alpha+\delta, \alpha+\gamma, \beta+\delta, \beta+\gamma\} $, which leads to the upper bound of the objective function in (\ref{eq:obj}) as $\sum_{i=1}^6\lambda_i$. Finally, we have an upper bound of the sum-DoF as
\begin{align}
\sum_{k=2}^{K}d_{1,k}+\sum_{k=2}^{K}d_{k,1}  \leq &\max_{\lambda_1,\ldots,\lambda_6} \sum_{i=1}^6\lambda_i = 1,
\end{align}
where the last equality follows from the fact  $\sum_{i=1}^{6}\lambda_i=1$.

By symmetry, the sum-DoF bounds are the same for the other user cooperation scenarios. This completes the proof.

 \section{ Another Transmission Scheme of Example 1}
We provide a different scheme with Example 1, which shows the same achievability of $\frac{4}{3}$ DoF for the two-pair two-way relay channel when $N=2$.

In the first time slot, user 1 and user 3 send signals $x_1[1]=s_{3,1}$ and $x_3[1]=s_{1,3}$. Due to half-duplex constraint, the received signals at user 2, user 4, and the relay are given by
\begin{eqnarray}
{ y}_{2}[1] &=&{h}_{2,1}[1]s_{3,1}+  {h}_{2,3}[1]s_{1,3}, \nonumber \\
{ y}_{4}[1] &=& {h}_{4,1}[1]s_{3,1}+  {h}_{4,3}[1]s_{1,3}, \nonumber \\
{\bf y}_{\textrm{R}}[1]&=&{\bf h}_{\textrm{R},1}[1]s_{3,1} +{\bf h}_{\textrm{R},3}[1] s_{1,3}.\label{eq:apen3}
\end{eqnarray}
Using two antennas at the relay, it decodes the transmitted data symbols $s_{3,1}$ and $s_{1,3}$.

In time slot 2, user $2$ and user $4$ send signals ${x}_{2}[2]={ s}_{4,2}$ and ${ x}_{4}[2]={s}_{{2,4}}$ and user 1, user 3, and the relay receive
\begin{eqnarray}
y_1[2]&=&{ h}_{1,2}[2]{ s}_{4,2}+{ h}_{1,4}[2]{ s}_{{2,4}} \nonumber \\
y_3[2]&=&{ h}_{3,2}[2]{ s}_{{4,2}}+{ h}_{3,4}[2]{ s}_{{2,4}}\nonumber \\
{\bf y}_{\textrm{R}}[2]&=&{\bf { h}}_{\textrm{R},2}[2]{ s}_{4,2} +{\bf { h}}_{\textrm{R},4}[2] { s}_{{2,4}}.
\end{eqnarray}
From the second time slot transmission, user 1 and user 3 obtain a linear equation consisted of interference symbols, while the relay decodes ${ s}_{{4,2}} $ and ${ s}_{2,4} $ by using a ZF decoder.

 In the third time slot, the relay transmits signal,
 \begin{eqnarray}
{\bf x}_R[3]={\bf v}_{3,1}[3]s_{3,1}+{\bf {v}}_{1,3}[3]{ s}_{1,3}+{\bf v}_{4,2}[3]s_{4,2}+{\bf {v}}_{2,4}[3]{ s}_{2,4}.
 \end{eqnarray}
To eliminate interference signals at the users based on the observed interference signals during time slot 1 and 2, the relay precoding vectors are designed so that the following space-time interference alignment conditions are satisfied,
\begin{eqnarray}
\left[%
\begin{array}{cc}
{\bf h}_{2,\textrm{R}}^*[3] & {\bf h}_{4,\textrm{R}}^*[3]
\end{array}%
\right]{\bf v}_{3,1}[3]=\left[%
\begin{array}{c}
  h_{2,1}[1]\\
 h_{4,1}[1]\\
\end{array}%
\right], \\
\left[%
\begin{array}{cc}
{\bf h}_{2,\textrm{R}}^*[3] & {\bf h}_{4,\textrm{R}}^*[3]
\end{array}%
\right]{\bf v}_{1,3}[3]=\left[%
\begin{array}{c}
  h_{2,3}[1]\\
  h_{4,3}[1]\\
\end{array}%
\right], \\
\left[%
\begin{array}{cc}
{\bf h}_{1,\textrm{R}}^*[3] & {\bf h}_{3,\textrm{R}}^*[3]
\end{array}%
\right]{\bf v}_{4,2}[3]=\left[%
\begin{array}{c}
  h_{1,2}[2]\\
  h_{3,2}[2]\\
\end{array}%
\right], \\\left[%
\begin{array}{cc}
{\bf h}_{1,\textrm{R}}^*[3] & {\bf h}_{3,\textrm{R}}^*[3]
\end{array}%
\right]{\bf v}_{2,4}[3]=\left[%
\begin{array}{c}
  h_{1,4}[2]\\
  h_{3,4}[2]\\
\end{array}%
\right].
 \end{eqnarray}
Since all channel elements are i.i.d., four precoding vectors can be obtained with high probability.

Thus, the received signals at user 1 for time slot 3 is given by
\begin{eqnarray}
{y}_{1}[3]&=&{\bf h}^*_{1,\textrm{R}}[3]{\bf x}_{\textrm{R}}[3] \nonumber \\
&=&{\bf h}^*_{1,\textrm{R}}[3]{\bf v}_{1,3}[3]s_{1,3}\!+\!\underbrace{{\bf h}^*_{1,\textrm{R}}[3]{\bf { v}}_{3,1}[3]{ s}_{3,1}}_{\textrm{Self-interference}}+\underbrace{{\bf h}^*_{1,\textrm{R}}[3]{\bf {v}}_{4,2}[3]{s}_{4,2}+ {\bf h}^*_{1,\textrm{R}}[3]{\bf {v}}_{2,4}[3]{s}_{2,4}}_{y_1[2]=h_{1,2}[2]s_{4,2}+h_{1,4}[2]s_{2,4}}.
\end{eqnarray}
By using successive interference cancellation, user 1 obtains the desired data symbol ${ s}_{{1,3}}$ as
\begin{eqnarray}
{y}_{1}[3]-y_1[2]-{\bf h}^*_{1,\textrm{R}}[3]{\bf { v}}_{3,1}[3]{ s}_{3,1}={\bf h}^*_{1,\textrm{R}}[3]{\bf v}_{1,3}[3]s_{1,3}.
\end{eqnarray}
By symmetry, the other three users obtain one desired signal each, which leads to achieve the $\frac{4}{3}$ of DoF. The main difference with the proposed scheme explained in Example 1 is that the relay needs current CSI knowledge of ${\bf h}^*_{\ell,\textrm{R}}[3]$ for $\ell\in\{1,2,3,4\}$ as well as outdated CSI $h_{i,j}[t]$ for $i,j,t\in\{1,2\}$.


%


\end{document}